\documentclass[journal,10pt,twocolumn,final]{IEEEtran}

\usepackage{amsmath,color,url}
\let\theoremstyle\relax
\usepackage{amsthm}
\usepackage{amssymb}
\usepackage[english]{babel}
\usepackage{graphicx,epstopdf,tikz,float}

\usetikzlibrary{arrows,positioning,calc}
\theoremstyle{plain}

\newtheorem{lemma}{Lemma}
\newtheorem{proposition}{Proposition}

\newtheorem{theorem}{Theorem}
\newtheorem{definition}{Definition}

\newtheorem{assumption}{Assumption}

\theoremstyle{remark}
\newtheorem{example}{Example}
\newtheorem{remark}{Remark}

\newcommand{\R}{\mathbb R}
\newcommand{\Z}{\mathbb Z}
\newcommand{\N}{\mathbb N}

\newcommand{\D}{\mathbb D}

\newcommand{\Rplus}{\R_+}

\newcommand{\sprod}[2]{\langle #1,#2\rangle} 

\newcommand{\cB}{\mathcal B}
\newcommand{\cC}{\mathcal C}

\newcommand{\cH}{\mathcal H}

\newcommand{\cJ}{\mathcal J}
\newcommand{\cK}{\mathcal K}
\newcommand{\cL}{\mathcal L}
\newcommand{\cM}{\mathcal M}

\newcommand{\cT}{\mathcal T}
\newcommand{\cU}{\mathcal U}

\newcommand{\cX}{\mathcal X}

\newcommand{\cZ}{\mathcal Z}

\newcommand{\rC}{\mathrm{C}}
\newcommand{\rD}{\mathrm{D}}

\newcommand{\rX}{\mathrm{X}}

\newcommand{\rZ}{\mathrm{Z}}

\newcommand{\x}{\times}
\newcommand{\st}{\,\mid\,}

\newcommand{\sr}{^\star}

\newcommand{\xb}{\mathbf{x}}

\newcommand{\vhi}{\varphi}

\newcommand{\scalf}{\varphi}
\newcommand{\wavf}{\psi}
\newcommand{\Scalf}{\Phi}
\newcommand{\Wavf}{\Psi}

\DeclareMathOperator{\col}{col}
\DeclareMathOperator{\img}{Im}
\DeclareMathOperator{\diag}{diag}
\DeclareMathOperator{\linspan}{span}
\DeclareMathOperator{\esssup}{ess.sup}
\DeclareMathOperator{\dom}{dom}

\DeclareMathOperator{\supp}{supp}
\DeclareMathOperator{\card}{card}
\DeclareMathOperator{\msv}{msv}
\DeclareMathOperator{\sat}{sat}

\newcommand{\cspan}{\overline{\linspan}}

\newcommand{\inv}{^{-1}}
\newcommand{\pinv}{^\dagger}
\newcommand{\argmin}[1]{\underset{#1}{\arg\min}\,}

\newcommand{\bV}{\mathbf{V}}

\newcommand{\bk}{\mathbf{k}}
\newcommand{\bK}{\mathbf{K}}
\newcommand{\bH}{\mathbf{H}}

\newcommand{\thmap}{\gamma}

\newcommand{\e}{{\rm e}}

\newcommand{\vep}{\varepsilon}
\DeclareMathOperator{\errObs}{oe}
\DeclareMathOperator{\errEst}{ee}

\newcommand{\PP}{{\rm P}}
\newcommand{\QQ}{{\rm Q}}

\newcommand{\dy}{{p}}
\newcommand{\dth}{{n_{\theta}}}
\newcommand{\ddth}{n_\theta}
\newcommand{\dx}{{n\dy}}

 \newcommand{\uT}{\underline{\rm T}}
 \newcommand{\pT}{\overline{\rm T}}
 
 \newcommand{\ain}{\alpha_{\rm in}}
 \newcommand{\aout}{\alpha_{\rm out}}
 
 \newcommand{\dg}{{\rm d}_g}
 
 \newcommand{\SPD}{\mathbb{SPD}}
 \newcommand{\CC}{\mathbb{C}_c(\R^n)}
 \newcommand{\CCR}{\mathbb{C}_c(\R)}
 
 \newcommand{\bh}{\mathbf{h}}
 
\DeclareMathOperator{\jumps}{Jmp}
\DeclareMathOperator{\flows}{Flw}
 
\newcommand{\costf}{{\rm k}}
\newcommand{\regf}{{\varrho}}
 
\DeclareMathOperator{\optmap}{Opt}

\newcommand{\din}{\delta_{\rm in}}
\newcommand{\dout}{\delta_{\rm out}}

\newcommand{\eout}{{e}_{\rm out}}

\allowdisplaybreaks

\def\middlebreak {\nulldelimiterspace0pt
	\allowbreak\mskip 0mu plus .5mu \nulldelimiterspace0pt}%

\begin{document}

	\onecolumn 
	\vspace{4em}
	\begin{quote}
		This is the post peer-review accepted version of: M. Bin and L. Marconi, ``Model Identification and Adaptive State Observation for a Class of Nonlinear Systems,'' accepted for publication in IEEE Transaction on Automatic Control (DOI: 10.1109/TAC.2020.3041238).  
		The published version is available online at  \url{https://ieeexplore.ieee.org/document/9272829}.
	\end{quote}
	
	\vspace{5em}
	\begin{quote}
		\emph{\textcopyright{}~2020 IEEE.  Personal use of this material is permitted.  Permission from IEEE must be obtained for all other uses, in any current or future media, including reprinting/republishing this material for advertising or promotional purposes, creating new collective works, for resale or redistribution to servers or lists, or reuse of any copyrighted component of this work in other works.}
	\end{quote}

	\twocolumn
	
	\title{Model Identification and Adaptive State Observation for a Class of Nonlinear Systems} %

\author{Michelangelo~Bin~and~Lorenzo Marconi%
\thanks{Michelangelo Bin is with the Department of Electrical and Electronic Engineering, Imperial College London, London, UK. Lorenzo Marconi is with the Department of Electrical, Electronic, and Information Engineering, University of Bologna, Bologna, Italy.}
}

\maketitle
\thispagestyle{empty}
\pagestyle{empty}
 

	\begin{abstract}
		In this paper we consider the joint problems of state estimation and model identification for a class of continuous-time nonlinear systems in output-feedback canonical form. An adaptive  observer is proposed that combines an extended high-gain observer and a discrete-time identifier. The extended observer provides the identifier with a data set permitting the identification of the system model and the identifier adapts the extended observer according to the new estimated model. The design of the identifier is approached as a system identification problem and sufficient conditions are presented that, if satisfied, allow different identification algorithms to be used for the adaptation phase. The cases of recursive least-squares and  multiresolution black-box identification via wavelet-based identifiers are specifically addressed. Stability results are provided relating the asymptotic estimation error to the prediction capabilities of the identifier. Robustness with respect to additive disturbances affecting the system equations and measurements is also established in terms of an input-to-state stability property relative to the noiseless estimates. 
	\end{abstract}
		\begin{IEEEkeywords}   
			Adaptive Observers, Identification for Control, High Gain Observers, Least Squares, Wavelets              
		\end{IEEEkeywords} 

\section{Introduction}
\subsection{Problem Description and Literature Overview}
We consider nonlinear systems of the form
\begin{equation}\label{s:x}
\begin{array}{lclr}
\dot x_i&=&x_{i+1} +d_i &\quad i=1,\dots,n-1\\
\dot x_n&=& \phi(x)+d_n\\
y&=&x_1 + \nu
\end{array}
\end{equation}
where $x=(x_1,\dots,x_n)\in\R^{\dx}$, $x_i\in\R^\dy$, is the state, $y\in\R^\dy$ is the measured output, $d:=\col(d_1,\dots,d_n) \in\R^{\dx}$ and $\nu\in\R^\dy$ are unknown disturbances, and $\phi:\R^{\dx}\to\R^\dy$ is a function  unknown to the designer and fulfilling technical assumptions specified later. For the class of systems \eqref{s:x}, in this paper we consider the  problem of designing an adaptive observer that, when the disturbances $d$ and $\nu$ are not present,  produces  asymptotically ``good'' estimates of both the system state $x$ and the system model $\phi$, and that, when instead $d$ and $\nu$ are present, guarantees  a corresponding ``input-to-state stability'' property relative to such asymptotic estimates. An adaptive observer with those properties is called in \cite{Marino2001} a \emph{robust adaptive observer}.  

State estimation is a problem of primary interest in control, and its applications are ubiquitous in all the related engineering areas. Having available good models, on the other hand, is of crucial importance in many contexts, as for instance in \emph{model predictive control} \cite{Grune2017} and control in presence of delays \cite{Krstic2009}, where models are used to cast predictions, in \emph{tracking} and \emph{output regulation} \cite{IsidoriBookNew}, in which they are used to model the exogenous references or disturbances acting on the system, and in signal processing and the related applications \cite{Kay1993,Cichocki2002}, in which models are used to extract information about the surrounding environment and to detect events. In this paper we jointly consider both the problems of estimating the system state $x$  and     its  model $\phi$.

The problem of designing adaptive observers for uncertain nonlinear systems boasts decades of active research, and it  constitutes nowadays an important branch of adaptive control.  The  class of systems \eqref{s:x}, in turn, is among the most studied, and the many contributions developed in the years mainly differ in terms of the hypotheses on the structure of the uncertain model here represented by $\phi$. Systems in which the uncertainty is concentrated in a set of parameters entering linearly in the model have been considered for instance in \cite{Bastin1988,Marino1992,Marino1995} and \cite{Marino2001}, where in the latter work also robustness with respect to additive disturbances (modeled by $d$ in \eqref{s:x}) is proved.  Related extensions to multivariable systems and more general forms appeared in \cite{Besancon2000,Zhang2002} in a disturbance-free setting.  Adaptive observers for models admitting  nonlinear parametrizations in the uncertain parameters started to appear in more recent times (see \cite{Farza2009,Tyukin2013,Afri2017,Besancon2017} and the references therein). In particular, in \cite{Farza2009} a general class of high-gain observers is endowed with an adaptation mechanism of the kind of that proposed in \cite{Zhang2002}. In \cite{Tyukin2013}, a fairly more general class of nonlinear parametrizations is considered for uniformly observable systems, where also robustness with respect to disturbances $d$ affecting the system dynamics is shown to hold. In \cite{Afri2017}, the theory of nonlinear Luenberger observers \cite{Andrieu2006} is applied to estimate the state and parameters of uncertain linear systems and, in \cite{Besancon2017}, more general systems exhibiting nonlinearities in both  the states   and the parameters are dealt with by using the same arguments of \cite{Besancon2007} in dealing with non-uniformly observable systems. 
 
Most of the existing approaches are strongly based on a canonical ``adaptive control'' perspective, in which all the uncertainty is concentrated in a  set of parameters with known dimension, whose knowledge would result in the knowledge of the \textit{true system} to be observed. In line with the certainty equivalence principle, uncertainty is usually dealt with by using an estimate of the true uncertain parameter, whose adaptation is carried out by ad hoc adaptation laws induced by Lyapunov analysis or  immersion arguments. While for linear systems parametric perturbations do actually exhaust the kind of model uncertainties we may care about, this is definitely not the case for nonlinear systems, where limiting to parametric perturbations means considering variations of functions in quite ``restricting'' topologies \cite{Bin2018CDC}. Furthermore, the assumption of a known parametrization is also \emph{not robust} relative to quite standard topologies,
in the sense that even if a nominal model admits the supposed parametrization, any arbitrarily ``small'' neighborhood of the nominal model will contain functions not having such a structure. This, indeed, makes the approaches constructed around the notion of a \emph{``true model''} conceptually fragile, especially in view of the fact that robustness to  disturbances is usually not shown. In particular, among the aforementioned approaches, only~\cite{Marino2001},~\cite{Tyukin2013} and~\cite{Afri2017} consider robustness to disturbances affecting the system dynamics ($d$ in \eqref{s:x}), and only  \cite{Afri2017} considers also measurement noise ($\nu$ in \eqref{s:x}). The result of \cite{Afri2017}, however, limits to linear systems, and the  robustness property is in general only \emph{local} in both $d$ and $\nu$, and can be made global  only under some additional boundedness conditions on the parametrization.

\subsection{Objectives, Results and Organization of the Paper}
The main goal of this paper is to propose a class of \emph{robust adaptive observers} estimating both the state and the model of system \eqref{s:x}, and whose construction is not tailored around the assumption of a known structure of the system's model $\phi$ and   does not necessarily rely on the existence of a \emph{``true model''} and of a corresponding \emph{``true parameter''} to estimate. Instead, we aim to approach model estimation as a \emph{system identification} problem \cite{Ljung1999}, with the goal of connecting the quality of the state and model estimation to the prediction performance of the employed identification scheme. The quest of estimating the true parameter is thus substituted with that of finding the \emph{best} model possible and, accordingly, the objective of asymptotic state estimation  is substituted by a weaker  \emph{optimality} property of the identified model, more likely achievable robustly. In this way, we aim at opening the doors  to nonlinear \emph{black-box} techniques and \emph{universal approximators} \cite{Sjoberg1995,Juditsky1995},  by adopting, however, standard analysis tools typical of systems and control theory.  

System identification is a very rich and developed research area, and an enormous variety of techniques and algorithms exist whose applicability strongly depends on the kind of signals under concern. Consistently with the aforementioned objective, we do not intend here to focus on a single identification algorithm, that would inexorably make sense only for a restricted class of models of $\phi$. Rather we aim to extract and provide generic sufficient conditions that an arbitrary identification algorithm must satisfy to be used in the framework, while leaving to the user the choice of the particular strategy to employ depending on the a priori qualitative information available on the system.  In these terms, we seek a solution which is  \emph{modular} in the choice of the adaptation strategy and, thus, which can be adapted to the particular application context and to the amount and quality of the a priori information that the designer has on the system.

To this end, we endow an extended high-gain observer with a generic discrete-time identifier satisfying some steady-state optimality and strong stability assumptions detailed later. The extended observer provides the identifier with a ``dirty'' data set that the identifier can use to extract information about the system model $\phi$. The identifier adapts the extended observer according to its best guess of $\phi$, thus   increasing the observer performance.   The main result of the paper is an \emph{approximate} estimation result stating that the adaptive observer produces an estimate of the system state and model that is asymptotically bounded by the ``size'' of the disturbances $d$ and $\nu$, and by the \emph{prediction capabilities} of the chosen identifier. 

The paper is organized as follows. In Section \ref{sec:observer} we present the adaptive observer; in Section \ref{sec:main} we give the main result; in Section \ref{sec:identifiers}, we show how an identifier implementing the well-known recursive least-squares scheme can be constructed, and based on that, we propose a recursive identifier performing a black-box identification procedure based on wavelet and multiresolution analysis. Finally in Section \ref{sec:example} we present some examples.

\subsection{Contribution of the Paper }
State estimation for systems of the form \eqref{s:x} in presence of uncertain $\phi$, alone, is a problem that does not actually need adaptation to give acceptable results when single experiments are concerned. As a matter of fact, in a noise-free context, standard high-gain observers can be used to obtain a \emph{practical} state estimate without knowing $\phi$ exactly \cite{KhalilNL}, or sliding-mode and homogeneous observers can be used to get a theoretically asymptotic estimate by just knowing a bound on~$\phi(x)$~\cite{Levant1998,Andrieu2008}. The use of adaptation, in turn, is typically motivated when also the model of the observed system is needed other than its state, or when one is interested to obtain an observer that is actually able to generate the trajectories of the observed system. The main contribution of this paper concerns this latter cases, where it intersects the already vast literature on adaptive observation. Compared to the existing approaches, in this paper adaptation is treated in more general terms as a system identification problem, and the proposed solution is \emph{modular} in the choice of the adaptation strategy, thus allowing the user to employ different  techniques  depending on the actual needs. This, in turn, requires a different approach to the analysis and synthesis that  is based on a logical separation of the roles of the observer and the identifier. 
  Furthermore, the proposed design is intrinsically \emph{robust} to measurement noise and additive disturbances, without the need of introducing tedious modifications of the adaptive strategy or the observer. 

Regarding the actual design of identifiers, as a particular case we show that the well-known least-squares schemes fit into the framework, and we propose a novel black-box identifier performing a wavelet multiresolution approximation of arbitrary $L^2$ functions. 

The observer is specifically designed to  support discrete-time identifiers. This is motivated by the fact that system identification approaches are usually discrete-time. Their use, however, comes with the price of a considerable technical overhead, requiring to study the overall interconnection in the formal context of \emph{hybrid dynamical systems} \cite{Goebel2012book}.

\subsection{Notation}
We denote by $\R$, $\N$ and $\Z$ the set of real, natural and integer numbers, and we let $\Rplus:=[0,\infty)$. If $S$ is a set, $\card{S}$ denotes its cardinality. We denote by $|\cdot|$ any norm whenever the underlying normed space is clear. If $S$ a subset of a normed vector space $\cX$ and $x\in\cX$, we let $|x|_S:=\inf_{s\in S}|x-s|$. With $A_1,\dots, A_n$ matrices, we denote by $\col(A_1,\dots, A_n)$ and $\diag(A_1,\dots, A_n)$ their column and diagonal concatenations. For an indexed family of matrices $\{A_b\}_{b\in\cB}$, we also use the notation $\col(A_b\st b\in \cB)$. We denote by $\msv(A)$ the minimum non-zero singular value of the matrix $A$. 
A function $\gamma:\Rplus\to\Rplus$ belongs to {\em class-$\cK$} ($\gamma\in\cK$) if it is continuous, strictly increasing and $\gamma(0)=0$. If moreover $\gamma(s)\to_{s\to\infty}\infty$, $\gamma$ is said to belong to {\em class-$\cK_\infty$} ($\gamma\in\cK_\infty$). A continuous function $\beta:\Rplus\x\Rplus\to\Rplus$ belongs to {\em class-$\cK\cL$} ($\beta\in\cK\cL$)  if for each $t\in\Rplus$,  $\beta(\cdot,t)\in\cK$ and for each $s\in\Rplus$, $\beta(s,\cdot)$ is decreasing and $\beta(s,t)\to_{t\to \infty} 0$. For $k\in\N$, $\cC^k$ denotes the set of $k$-times continuously differentiable functions. If $\phi$ is a function, $\supp\phi$ denotes its support. $L^2(\cX)$ denotes the Lebesgue space of square integrable functions $\cX\to\R$.  We denote by $\sprod{\cdot}{\cdot}$ the usual scalar product on $L^2(\cX)$ and we equip $L^2(\cX)$ with the  norm $|\cdot|:=\sqrt{\sprod{\cdot}{\cdot}}$.

In this paper we deal with \emph{hybrid systems}, which are dynamical systems whose state may exhibit both a continuous-time evolution and impulsive changes. Hybrid systems are formally described by equations of the form \cite{Goebel2012book}
\begin{equation}\label{s:hs0}
\Sigma: \left\{ \begin{array}{lclrl}
\dot x &= & F(x,u) && (x,u)\in  C \\
x^+ &= & G(x,u) && (x,u)\in D
\end{array}\right.
\end{equation}
where $x$ denotes the state, $u$ an exogenous input, and
$C$ and $D$ denote the sets in which continuous and discrete-time dynamics are allowed. In particular, the first equation of \eqref{s:hs0} means that the state can evolve according to $\dot x= F(x,u)$ whenever $(x,u)\in C$. The second equation means that the state can exhibit an impulsive change, from $x$ to $G(x,u)$, whenever $(x,u)\in D$.
The solutions of \eqref{s:hs0} are defined over \emph{hybrid time domains}, which generalize the usual time domains~$\Rplus$ and $\N$ of continuous and discrete time systems. In particular,  a \emph{compact hybrid time domain} is a subset of $\Rplus\x\N$ of the form $\cT=\cup_{j=0}^{J-1} [t_j,t_{j+1}]\x\{j\}$ for some  $J\in\N$ and $0=t_0\le t_1\le\dots\le t_J$. A set $\cT\subset\Rplus\x\N$ is called a \emph{hybrid time domain} if for each $(T,J)\in\Rplus\x\N$, $\cT\cap[0,T]\x\{1,\dots,J\}$ is a compact hybrid time domain. 
For $(t,j),(s,i)\in \cT$, we write $(t,j)\preceq (s,i)$ if $t+j\le s+i$,    we let   $t^j=\sup_{t\in\R}(t,j)\in\cT$, $t_j:=\inf_{t\in\R}(t,j)\in\cT$ and we define $j_t$ and $j^t$ similarly.  

A function $x:\cT\to\cX$ defined on a hybrid time domain $\cT$ is called a \textit{hybrid arc} if $x(\cdot,j)$ is locally absolutely continuous for each $j$. Hybrid arcs define the space in which the solutions of \eqref{s:hs0} live, and they play the same role of absolutely continuous signals  in a standard continuous-time setting. Further restrictions are needed for a hybrid arc to be used as input in~\eqref{s:hs0}. In particular, a  \emph{hybrid input} $u$ is a hybrid arc such that $u(\cdot,j)$ is locally essentially bounded and Lebesgue measurable for each $j$. A pair $(x,u)$ is called a \emph{solution pair} to \eqref{s:hs0} if  $x$ is a hybrid arc, $u$ is a hybrid input, and $(x,u)$ satisfy \eqref{s:hs0}.  We call a solution pair   {\em complete} if its time domain is unbounded.  We let $\dom x$ denote the domain of $x$, $\flows x\subset\Rplus$ the set of $t$ such that $(t,j)\in\dom x$ for some $j\in\N$, and $\jumps x\subset\N$ the set of $j$ such that $(t,j)\in\dom x$ for some $t\in\R$. In order to simplify the notation, we omit the jump (resp. flow) equation when the considered system has only continuous-time (resp. discrete-time) dynamics. If $x$ is constant during flows (resp. jumps), we neglect the ``$t$'' (resp ``$j$'') argument and we write $x(j)$ (resp $x(t)$), which we identify with the map $\jumps x \ni  j\mapsto x(t_j,j)$ (resp $\flows x\ni t\mapsto x(t,j_t)$). 
We say that a hybrid arc $x$ is \emph{eventually} in a set $X$ if for some $s\in\dom x$, $x(t,j)\in X$ for all $(t,j)\succeq s$.
With $u:\dom u\to\cU$ a hybrid input, and   $\Gamma(u):=\{(t,j)\in \dom u\st (t,j+1)\in \dom u\}$, for   $(t,j)\in\dom u$ we let $|u|_{(t,j)}:=\max\{   \sup_{(t,j)\in\Gamma(u),\,(0,0)\preceq(s,i)\preceq(t,j)} |u(s,i)|, \middlebreak \esssup_{(s,i)\in\dom u/\Gamma(u),(0,0)\preceq (s,i)\preceq (t,j)} |u(s,i)|\}$. If $u$ is constant during jumps (resp. flows), we  write $|u|_t$ (resp $|u|_j$) as short for $|u|_{(t,j)}$.
 We also let $|u|_{A,(t,j)}:=\big||u|_A \big|_{(t,j)}$ and $|u|_\infty:=\limsup_{t+j\to\infty,(t,j)\in\dom u} |u|_{(t,j)}$.
 Given a hybrid arc $x$ we will abbreviate $\limsup_{t+j\to\infty,(t,j)\in\dom x}x(t,j)$ with $\limsup_{t+j\to\infty} x$.  A collection of hybrid arcs $\cH$ is  ``eventually equibounded'' if there exist  $\tau,m>0$ such that $|x(t,j)|\le m$ for each $x\in \cH$ and $(t,j)\in\dom x$ satisfying $t+j\ge\tau$. The abbreviation ``ISS'' stands for ``input-to-state stability''.

\section{The Adaptive Observer}\label{sec:observer}
System \eqref{s:x} can be written in compact form as
\begin{equation}\label{s:xx}
\begin{array}{lcl}
\dot x &=& Ax+B \phi(x) + d \\
y&=& Cx+\nu 
\end{array}
\end{equation}
with 
\begin{align*}
A&: =\begin{pmatrix}
0_{(\dx-\dy)\x\dy}  & I_{\dx-\dy}\\
0_\dy & 0_{\dy\x(\dx-\dy)}
\end{pmatrix}  , & B&:=\begin{pmatrix}
0_{(\dx-\dy)\x\dy}\\ I_\dy
\end{pmatrix}\\
C&:=\begin{pmatrix}
I_\dy & 0_{\dy \x(\dx-\dy)}
\end{pmatrix}.
\end{align*}
On system \eqref{s:xx} we make the following  assumption.
\begin{assumption}\label{ass:x} The function $\phi$ is locally Lipschitz. Moreover,~$d$ and $\nu$ are hybrid inputs and  there exist (known)  compact sets $\cX_0\subset\cX\subset\R^\dx$ and, for each $x_0\in\cX_0$, a set $\D(x_0)$ of   hybrid inputs with values in $\R^\dx$, such that each solution pair $(x,d)$ to \eqref{s:xx} with $x(0)\in\cX_0$ and $d\in\D(x(0))$  satisfies $x(t)\in\cX$ for all $t\in\dom x$.
\end{assumption}
\begin{remark}
	Assumption~\ref{ass:x} is a boundedness condition requiring   the solutions to \eqref{s:xx} originating inside a known, arbitrary  compact set $\cX_0$ to be uniformly bounded. Since the effect of $d$ on the time evolution of $x$ depends on the value of $\phi$, then Assumption \ref{ass:x} imposes a \emph{joint} restriction  on   $\phi$ and the class of admissible disturbances $d$. 
\end{remark}
\begin{remark}
	We underline that the exogenous signals $d$ and~$\nu$ are required to be \emph{hybrid inputs} and, as such,  they must be locally essentially bounded and Lebesgue measurable (see the notation section). Therefore, the results presented in this paper \emph{do not} completely cover the case in which $d$ and $\nu$ are  realizations of common stochastic processes such as Gaussian processes, as they may not be  locally essentially bounded, thus failing to be hybrid inputs. In turn, the results presented in this paper only apply to those realizations that fit into the definition of hybrid input  and for which Assumption \ref{ass:x} above holds.
\end{remark}

The proposed adaptive observer is  a system whose dynamics is described by the following hybrid equations 
\begin{subequations}\label{s:obs}
\begin{equation}
\begin{aligned}
&\left\{ \begin{array}{lcl}
\dot\tau &=& 1\\
\dot{\hat x}  &=& A\hat x + B\xi +  \Lambda_1(g) (y-C\hat x)  \\
\dot\xi  &=& \psi(\theta,\hat x,\xi) + \Lambda_2(g) (y-C\hat x)\\
\dot z &=& 0
\end{array} \right.\\
& \qquad (\tau,\hat x,\xi,z,y)\in \rC_\tau \x \R^{\dx}\x\R^{\dy}\x\cZ\x\R^\dy\\
&\left\{ \begin{array}{lcl}
\tau^+ &=& 0\\
\hat x^+ &=& \hat x\\
\xi^+ &=& \xi \\
 z^+ &=& \vhi(z,\hat x,\xi )  
\end{array} \right.\\
& \qquad (\tau,\hat x,\xi,z,y)\in \rD_\tau \x \R^{\dx}\x\R^{\dy}\x\cZ\x \R^\dy
\end{aligned}
\end{equation}
with
\begin{equation}
	\theta  = \thmap(z) \ \in\R^{\dth},
\end{equation} 
\end{subequations}
and in which $\Lambda_1(g)\in\R^{\dx\x\dy}$, $\Lambda_2(g)\in\R^{\dy\x\dy}$, $\psi:\R^\dth\x\R^\dx\x\R^\dy\to\R^\dy$,   $\vhi:\cZ\x\R^\dx\x\R^\dy\to\cZ$, $\thmap:\cZ\to\R^\dth$, and
\begin{align}\label{d:Ct_Dt}
\rC_\tau&:=[0,\pT],&\rD_\tau&:=[\uT,\pT],
\end{align}
where $g>0$ is a control parameter, $\dth\in\N$, $\cZ$ is a finite-dimensional normed vector space, and $\uT,\pT\in\Rplus$ satisfy  $0<\uT\le\pT<\infty$.
The observer \eqref{s:obs} is composed of different subsystems. The subsystem $\tau$ is a \emph{clock}, whose tick determines the next jump time. As clear from the definition of the sets $\rC_\tau$ and $\rD_\tau$, the jumps of \eqref{s:obs} need not be periodic, and any  clock strategy is possible provided that the  time between two successive jumps is lower-bounded by $\uT$ and upper-bounded by $\pT$.
The subsystem formed by $\hat x:=\col(\hat x_1,\dots,\hat x_n)\in\R^{\dx}$ and $\xi\in\R^{\dy}$ is an \emph{extended  observer}, while the subsystem $z$, called the \emph{identifier},  is a dynamical system implementing the chosen identification scheme. The identifier is a discrete-time system  processing the state $(\hat x,\xi)$ of the extended observer and giving as output  the variable $\theta\in\R^{\dth}$, through which it affects the observer's dynamics by adapting the ``consistency term'' $\psi(\theta,\hat x,\xi)$.  The construction of the identifier and the extended observer  subsystem is detailed in the next sections.

\subsection{The Identifier}
The identifier is a discrete-time system constructed to solve an identification problem cast on the samples of the following system, referred to as the \emph{core process}.
\begin{equation}\label{s:core}
\begin{array}{lcl}
 \begin{aligned}
&\left\{\begin{array}{lcl}
\dot\tau &=& 1\\
\dot x &=& Ax+B\phi(x) + d
\end{array} \right.   &   (\tau,x,d)&\in\rC_\tau\x\cX\x\R^\dx\\
&\left\{\begin{array}{lcl}
\tau^+&=& 0\\
 x^+ &=& x
 \end{array}\right.  &  (\tau,x,d)&\in\rD_\tau\x\cX\x\R^\dx
\end{aligned}
\end{array}
\end{equation}
with outputs $\ain\sr\in\R^\dx$ and $\aout\sr\in\R^\dy$ defined as
\begin{align*}
\ain\sr &:= x,& \aout\sr &:=\phi(x),
\end{align*}
and with initial conditions $(\tau(0),x(0))$ belonging to $(\rC_\tau\cup\rD_\tau)\x\cX_0$ and disturbances $d$ restricted to $\D(x(0))$.
At each jump time $j$, the inputs $\ain\sr$ and $\aout\sr$ provide a data set $\{(\ain\sr(t^i),\aout\sr(t^i))\}_{i=0}^{j-1}\middlebreak=\{(x(t^i),\phi(x(t^i))\}_{i=0}^{j-1}$ from which the identifier can infer a guess of $\phi$. The input $\ain\sr=x$, indeed, plays the role of a \emph{regressor} (or independent variable), while $\aout\sr=\phi(x)=\phi(\ain\sr)$ that of the \emph{dependent variable}. The role of the identifier is then to find the  model relating $\ain\sr$ and $\aout\sr$ which fits at best  the samples contained in the data set.

The core process plays a fundamental ``qualitative'' role in the definition of identification problem, as the a priori information that the designer has on its solutions determines the choice of a set $\cM$ (here called the  \emph{model set} \cite{Ljung1999}) of possible models $\hat\phi$ where to search for the best one  and, consequently, the choice of the identification algorithm.  As customary in system identification and for clear implementation constraints, we limit here to finite-dimensional model sets that, for some $\dth\in\N$, can be expressed as
\begin{equation*}
\cM:=\big\{ \hat\phi(\theta,\cdot): \R^\dx\to\R^\dy\st  \theta\in\R^\dth \big\}.
\end{equation*} 

The problem of selecting the model set is a rich area of system identification (see e.g. \cite{Ljung1999,Sjoberg1995,Juditsky1995}) and, depending on the quantity and quality of information the designer has on the core process, it may range from a family of functions $\hat\phi(\theta,\cdot)$ with a very specific form, such as linear regressions, to ``non-parametric'' models represented by \emph{universal approximators} such as \emph{wavelets} or \emph{neural networks} \cite{Juditsky1995}. 
In this phase we do not deal with the selection of the model set, which we assume to be fixed by the user, and we instead focus on the design of the identification algorithm. In particular, once $\cM$ is fixed, the problem of finding the ``best'' element $\hat\phi(\theta,\cdot)\in\cM$ can be formally cast by defining an optimization problem on the data set generated by the core process as follows. For each model $\hat\phi(\theta,\cdot)\in\cM$ and each solution pair $((\tau,x),d)$ to \eqref{s:core}, we define the \emph{prediction error}
\begin{equation}\label{d:pred_error}
\varepsilon(\theta,x) := \phi(x)-\hat\phi(\theta,x)= \aout\sr- \hat\phi(\theta, \ain\sr) .
\end{equation}
Then, with  each $j\in\jumps(\tau,x)$ we associate a cost functional $\cJ_{(\tau,x)}(j,\cdot):\R^\dth\to\Rplus$ having the generic expression
\begin{equation}\label{d:J}
\cJ_{(\tau,x)}(j,\theta) := \sum_{i=0}^{j-1}\costf\big(i,j,\varepsilon(\theta,x(t^i))\big)  + \regf(\theta),
\end{equation}
where $\costf:\N\x\N\x\R^\dy\to\Rplus$ and $\regf:\R^\dth\to\Rplus$ are user-decided continuous functions called respectively the \emph{integral cost} and \emph{regularization term}. The integral cost weights how a given $\theta$ fits previous samples of the data set, and its dependency on the current time $j$ and the past ones $i$ can be exploited to weight  more and less recent errors differently in the sum. Regularization is an important and widely spread practice in identification (see e.g. \cite{Sjoberg1993}), and the regularization term in \eqref{d:J} can be used to constraint the ``size'' of the parameter and to make ill-conditioned problems numerically treatable. For further details the reader is referred to \cite{Ljung1999,Sjoberg1993}.

For each solution pair $((\tau,x),d)$ to \eqref{s:core} and each $j\in\jumps(\tau,x)$, with  \eqref{d:J} we associate the set-valued map
\begin{equation*}
\optmap_{(\tau,x)}(j) := \argmin{\theta\in\R^\dth} \cJ_{(\tau,x)}(j,\theta),
\end{equation*}
which collects, at each $j$, the set of minima of $\cJ_{(\tau,x)}(j,\cdot)$.
The identifier subsystem  of \eqref{s:obs} is then a system designed to ``track'' the map of minima $\optmap_{(\tau,x)}$ when fed with the ideal input $(\ain\sr,\aout\sr)=(x,\phi(x))$. 

Clearly, the variables $\ain\sr$ and $\aout\sr$ are not available for feedback, as the only measured quantity is the system's output $y$. The identifier is thus fed in \eqref{s:obs} with the input $(\ain,\aout):=(\hat x,\xi)$, which provides a \emph{proxy} for the ideal input  $(\ain\sr,\aout\sr)$. This motivates asking a stronger property than ``just'' asymptotic tracking of $\optmap_{(\tau,x)}$, expressed in terms of an ISS requirement relative to disturbances possibly summing to $(\ain\sr,\aout\sr)$. 
More precisely, the required properties  are characterized in the forthcoming definition, by considering the cascade between the core process \eqref{s:core} and the discrete-time system
\begin{equation}\label{s.identifier}
\begin{array}{lcl}
z^+ &=& \vhi(z,\ain,\aout)\\
\theta  &=& \thmap(z)
\end{array}
\end{equation}
 obtained by letting $(\ain,\aout):=(\ain\sr,\aout\sr)+(\din,\dout)$, where $\delta:=(\din,\dout)$ is an exogenous disturbance, and reading as
\begin{equation}\label{s:core_identifier}
\begin{array}{lcl}
 \begin{aligned}
&\left\{\begin{array}{lcl}
\dot\tau &=& 1\\
\dot x&=& Ax+B\phi(x)+d\\
\dot z&=& 0
\end{array} \right.		\\&\qquad (\tau,x,z,d,\delta)\in\rC_\tau\x\cX\x\cZ\x\R^{\dx}\x\R^{\dx+\dy}\\
&\left\{\begin{array}{lcl}
\tau^+&=& 0\\
x^+ &=& x\\
z^+ &=& \vhi(z,\ain\sr+\din,\aout\sr+\dout) 
\end{array}\right. \\&\qquad (\tau,x,z,d,\delta)\in\rD_\tau\x\cX\x\cZ\x\R^{\dx}\x \R^{\dx+\dy}
\end{aligned}
\end{array}
\end{equation}
with output $\theta=\gamma(z)$.
\begin{definition}[Identifier Requirement]
	The tuple $(\cM,\cZ,\vhi,\thmap)$ is said to satisfy the \emph{Identifier Requirement} relative to $\cJ$ if there exist a compact set $\rZ\sr\subset\cZ$,  $\beta_z\in\cK\cL$, two Lipschitz functions $\rho_\theta,\,\rho_z\in\cK$, and for each solution pair $((\tau,x),d)$ to \eqref{s:core} with $x(0)\in\cX_0$ and $d\in\D(x(0))$, a hybrid arc $z\sr:\dom(\tau,x)\to\cZ$ and a $j\sr\in\N$, such that $((\tau,x,z\sr),(d,\delta))$ with $\delta=0$ is a solution pair to \eqref{s:core_identifier} satisfying $z\sr(j)\in\rZ\sr$ for all $j\ge j\sr$, and the following properties hold:
	\begin{enumerate}
		\item \textbf{Optimality:} For all  $j\ge j\sr$, the signal $\theta\sr:=\gamma(z\sr)$ satisfies
		\begin{equation*}
		 \theta\sr(j)\in\optmap_{(\tau,x)}(j) .
		\end{equation*}
		\item \textbf{Stability:} For every  hybrid input $\delta$, every solution pair to~\eqref{s:core_identifier} of the form $((\tau,x,z),(d,\delta))$, with $((\tau,x),d)$ the same as above, satisfies  for all $j\in\jumps (\tau,x,z)$
		\begin{equation*}
		\begin{aligned}
		|z(j)-z\sr(j)|&\le \max \big\{\beta_z(|z(0)-z\sr(0)|,j),\, \rho_{z}(|\delta|_j) 	\big\}.
		\end{aligned}
		\end{equation*} 
		
		\item \textbf{Regularity:} The map $\thmap$ satisfies
		\begin{equation} 
		|\thmap(z) -\thmap(z\sr) | \le \rho_\theta(|z -z\sr |)
		\end{equation}
		for all $(z,z\sr)\in\cZ\x\rZ\sr$ and, for each $\theta\in\R^\dth$, the map $\hat\phi(\theta,\cdot)$ is $\cC^1$ with locally Lipschitz derivative.
	\end{enumerate}

\end{definition}
The Identifier Requirement collects all the sufficient conditions that an identifier must satisfy to be embedded in the observer \eqref{s:obs}. In addition to regularity of the model $\hat\phi$ and the output map $\thmap$,  it requires the existence of a \emph{steady-state}~$z\sr$ for the identifier associated with the ideal input $(\ain\sr,\aout\sr)$ whose output $\theta\sr$ is \emph{optimal} relative to the cost functional \eqref{d:J}, and that an ISS property holds relative to such  optimal steady state and with respect to additive disturbances $\delta=(\din,\dout)$ affecting  $(\ain\sr,\aout\sr)$.

 Constructive design techniques implementing the relevant case of least-squares schemes and a wavelet-based black-box identifiers are postponed to Section \ref{sec:identifiers}.

\begin{remark}\label{rmk.EIV}
Since the disturbance $\delta=(\din,\dout)$ enters in both the   variables $\ain$ and  $\aout$, the identification problem underlying the Identifier Requirement fits into the context of \emph{Errors-In-Variables}  identification \cite{Soderstrom2007}. It is well-known (see e.g. \cite[Section 10]{Soderstrom2007}) that the presence of the disturbances in all the variables may cause a \emph{bias} in the parameter estimate even for ``simple'' least squares identifiers. Nevertheless, we remark that this is not in contrast with the Identifier Requirement as long as such bias vanishes continuously with the disturbances (see also Example~\ref{ex.noisy_regressor} in Section~\ref{sec:identifiers:ls} for a more detailed discussion in the least squares case).  In fact, when $z(0)=z\sr(0)$,  the combination of the stability and regularity properties of the Identifier Requirement implies $\sup_{j\ge j\sr} |\theta(j)-\theta\sr(j)| \le \rho_\theta(\rho_z( \sup_{j\ge j\sr} |\delta(j)|))$. Hence, the Identifier Requirement constraints the set of admissible identifiers to those  guaranteeing  that the uncertainty on the parameter estimate is bounded by a continuous image of the ``size'' of the  disturbances affecting the variables, whether such uncertainty comes from a bias or not.  We also remark that this is in line with typical results in the \emph{Set Membership} identification literature, in which the feasible set containing the true model is directly related to the maximum bound on the disturbances (see e.g. \cite{Milanese2004}).

Finally, we remark that the regularity property of the Identifier Requirement also implies that $\thmap$ is continuous on~$\rZ\sr$. Hence, $\gamma(\rZ\sr)$ is compact. This, in turn, yields an implicit assumption on the parameter $\theta\sr(j)$ to be identified, which is thus required to lie, eventually, in a known compact set (see also Example \ref{ex.noisy_regressor} in Section \ref{sec:identifiers:ls}).
\end{remark}

\subsection{The Extended  Observer}\label{sec:extObs}
The extended  observer subsystem $(\hat x,\xi)$ of \eqref{s:obs} follows a canonical high-gain construction \cite{KhalilNL}. In particular, we define the matrices $K_1,\dots,K_{n+1}\in\R^{\dy\x\dy}$  so that, with $K:=\col(K_1,\dots,K_{n+1})$, the matrix
\begin{equation}\label{d:M}
M:=\begin{pmatrix}
A & B\\
0 & 0
\end{pmatrix} - K\begin{pmatrix}
C & 0 
\end{pmatrix}
\end{equation}
is Hurwitz. In this respect, we notice that a feasible choice is to take $K_i:= k_i I_\dy$ with $k_i\in\R$ such that the polynomial $\lambda^{n+1}+k_1\lambda^{n}+\dots+k_{n}\lambda+ k_{n+1}$ has only roots with negative real part. With $g>0$ a control parameter to be fixed, we   let 
\begin{equation}\label{d.Lambdas}
\begin{aligned}
\Lambda_1(g) &:= \col(g K_1, g^2 K_2,\dots, g^n K_n), \\ \Lambda_2(g) &:= g^{n+1} K_{n+1}. 
\end{aligned}
\end{equation}
We then define arbitrarily two compact sets $X\sr\subset\R^\dx$ and $\Xi\sr \subset \R^\dy$ that are sets in which $x$ and $\phi(x)$  are supposed to lie eventually. Again, the choice of those sets is guided by the  qualitative and quantitative a priori information that the designer has on the system \eqref{s:x} and, under Assumption \ref{ass:x}, they are ideally taken so that $\cX\subset X\sr$ and $\phi(\cX)\subset\Xi\sr$. Moreover, with $\rZ\sr$ given by the Identifier Requirement, we let $\Theta\sr\subset\R^\dth$ be a compact set including $\thmap(\rZ\sr)$. We then choose $\psi$ as a bounded function obtained by ``saturating''  the function
\begin{equation*}  
 (\theta,\hat x,\xi ) \mapsto \dfrac{\partial \hat\phi (\theta,\hat x)}{\partial \hat x}\Big( A\hat x + B\xi   \Big)  
\end{equation*}
on the compact set $\Theta\sr\x X\sr\x\Xi\sr$ constructed above. More precisely,  we let $\psi$ be  a continuous function   satisfying 
\begin{equation}\label{d:psi_omega}
\begin{aligned}
\psi(\theta,\hat x,\xi) & = \dfrac{\partial \hat\phi (\theta,\hat x)}{\partial \hat x}\Big( A\hat x + B\xi   \Big) 
\end{aligned}
\end{equation}
for all $(\theta,\hat x,\xi)\in\Theta\sr\x X\sr\x \Xi\sr$    and
\begin{equation}\label{bound:psi_omega} 
|\psi(\theta,\hat x,\xi)| \le\bar\psi  \quad \forall(\theta,\hat x,\xi)\in\R^\dth\x\R^\dx\x\R^\dy, 
\end{equation}
 for some $\bar\psi\in\Rplus$ fulfilling
\begin{equation*}
\bar\psi > \max_{(\theta,\hat x,\xi)\in \Theta\sr\x X\sr\x\Xi\sr}\left|\dfrac{\partial \hat\phi (\theta,\hat x)}{\partial \hat x}\Big( A\hat x + B\xi   \Big) \right|,
\end{equation*}
 which exists by continuity, in view of the regularity item of the Identifier Requirement.

%
%
  
\section{Main Result}\label{sec:main}
 The interconnection between the system \eqref{s:x} (restricted on $\cX$) and the observer \eqref{s:obs} reads as follows
\begin{subequations}\label{s:x_obs}
\begin{equation}
\begin{aligned}
&\left\{ \begin{array}{lcl}
\dot\tau &=& 1\\
\dot x  &=& A x+B \phi(x) + d \\
\dot{\hat x}  &=& A \hat x  +  B\xi +\Lambda_1(g)(C(x-\hat x)+\nu)\\
\dot\xi  &=& \psi(\theta,\hat x,\xi) + \Lambda_2(g)(C(x-\hat x)+\nu)\\
\dot z &=& 0 
\end{array} \right.\\
& \qquad (\tau,x,\hat x,\xi,z,d,\nu)\in \rC \\
&\left\{ \begin{array}{lcl}
\tau^+ &=& 0\\
x^+&=&x,\quad 
\hat x^+  =  \hat x,\quad \xi^+   =  \xi \\
z^+ &=& \vhi(z,\hat x,\xi) 
\end{array} \right.\\
& \qquad (\tau,x,\hat x,\xi,z,d,\nu)\in \rD 
\end{aligned}
\end{equation}
with
\begin{equation}
	\theta=\gamma(z),
\end{equation}
\end{subequations}
and in which $\rC:=\rC_\tau \x \cX \x\R^{\dx}\x\R^{ \dy}\x\cZ\x\R^{\dx}\x\R^\dy$ and $\rD:=\rD_\tau \x \cX \x\R^{\dx}\x\R^{ \dy}\x\cZ\x\R^{\dx}\x\R^\dy$. We remark that restricting the flow and jump sets of $x$ to $\cX$  does not destroy completeness of  maximal solutions originating in~$\cX_0$ and with $d\in\D(x(0))$ in view of Assumption \ref{ass:x}.

In the following we will use the short notation  $\xb:=(\tau,x,\hat x,\xi,z)$. For each solution pair $(\xb,(d,\nu))$ to \eqref{s:x_obs}  with $x(0)\in\cX_0$ and  $d\in\D(x(0))$, $((\tau,x),d)$ is a solution pair to the core process \eqref{s:core}. We thus can associate with $(\xb,d)$ the hybrid arcs $z\sr$ and $\theta\sr$ produced by the Identifier Requirement for $((\tau,x),d)$. We then define the \emph{optimal prediction error} as
 \begin{equation*}
 \varepsilon\sr(t,j):= \varepsilon(\theta\sr(j),x(t)).
 \end{equation*}
For the sake of compactness, we also let
 \begin{equation}\label{d.oe_ee_dg}
 \begin{aligned}
 \errObs(\xb) &:= \max\{ |\hat x-x|,\, |\xi-\phi(x)|  \}\\
  \errEst(\xb,\theta\sr)  &:= \max\{ |x-\hat x|, \, |\theta-\theta\sr| \}\\
  \dg &:= (g\inv d_1, g^{-2} d_2,\dots, g^{-n} d_n).
 \end{aligned} 
 \end{equation}
Then the following theorem states the main result of the paper.
\begin{theorem}\label{thm:main}
Suppose that Assumption \ref{ass:x} holds,  and consider the observer \eqref{s:obs} constructed according to Section \ref{sec:observer}.  Then the following hold:
\begin{enumerate}
	\item There exist  $\rho_0,\rho_1,\rho_2,\rho_3>0$ such that every solution pair $(\xb,(d,\nu))$ to \eqref{s:x_obs} with $x(0)\in\cX_0$ and $d\in\D(x(0))$ satisfies for all $t\in\flows(\xb)$
	\begin{align*}
	&|x_i(t)-\hat x_i(t)| \le \max\Big\{ g^{i-1} \rho_0 \e^{-\rho_1gt}  \errObs(\xb(0)) , \\
	&\qquad\qquad\qquad\qquad\qquad\rho_2 g^{i-n-1},  \rho_3 g^{i-1} |(\dg,\nu)|_t  \Big\}  .
	\end{align*}

	\item If in addition $(\cM,\cZ,\vhi,\thmap)$ satisfies the Identifier Requirement relative to some cost functional $\cJ$, then there exist a locally Lipschitz $\rho\in\cK$ and $g \sr(\uT),\,r>0$ such that, if $g\ge g \sr(\uT)$, then every complete solution pair $(\xb,(d,\nu))$ to \eqref{s:x_obs} with $x(0)\in\cX_0$ and $d\in\D(x(0))$ for which $(x,\phi(x))$ is eventually in $X\sr\x\Xi\sr$ satisfies
	\begin{equation*}
	\begin{aligned}
	&\limsup_{t+j\to\infty} \errEst(\xb,\theta\sr) \le \rho\Big( \max\Big\{ \limsup_{t+j\to\infty} |\varepsilon\sr|,\\& \qquad\qquad\qquad\qquad\qquad\qquad g^n \limsup_{t\to\infty}|(\dg,\nu)| \Big\}\Big) \\
	&\limsup_{t\to\infty}|\hat x_i-x_i|\le \max\Big\{ g^{i-n} \limsup_{t+j\to\infty}|\vep\sr|,\\& \qquad\qquad\qquad\qquad\qquad\qquad r g^{i-1}\limsup_{t\to\infty}|(\dg,\nu)|  \Big\}.
	\end{aligned}
	\end{equation*} 
\end{enumerate} 
\end{theorem}

 Theorem \ref{thm:main} is proved in Appendix \ref{apd:proofThm}. The first claim recovers some well-known  properties of standard non-adaptive high-gain observers, showing that the introduction of adaptation preserves them. In particular, the claim establishes a practical ISS property of the estimate $\hat x-x$ relative to the disturbances~$(d,\nu)$, which boils down to a practical state estimation result when $(d,\nu)=0$. It also underlines how the effect of~$d$ and $\nu$ may be amplified by taking high values of~$g$, thus leading to a necessary compromise in its choice. 
 
 The second claim states that, if the identifier satisfies the requirement, and $(x,\phi(x))$ is eventually in the  set $\rX\sr\x\Xi\sr$, then the state and parameter estimation errors are asymptotically bounded by the optimal prediction error attained by the identifier, other than  the disturbances $d$ and $\nu$. We remark that the optimality item of the Identifier Requirement plays no role in the stability analysis yielding the claim of Theorem~\ref{thm:main}. It however guarantees that the target steady state $\theta\sr$ of  $\theta$ is optimal in the desired sense. Moreover, it implies that, whenever a \emph{``true model''} characterized by a null prediction error exists in the model set, then, in absence of regularization in \eqref{d:J}, the claim of the theorem can be strengthen to an ISS-like condition for the estimation errors with respect to $(d,\nu)$, which becomes asymptotic stability in a disturbance-free setting.

%
%
\section{On the Design of Identifiers}\label{sec:identifiers}
\subsection{Least-Squares Identifiers}\label{sec:identifiers:ls}
In this section we design an identifier satisfying the Identifier Requirement that implements a recursive least-squares identification scheme. For ease of notation  we limit to the single-variable case ($\dy=1$), and we remark that a multivariable identifier can be obtained by the composition of multiple single-variable identifiers. 

We take as model set $\cM$ the set of functions $\hat\phi$ of the form
\begin{equation*}
\hat\phi(\theta,\cdot) := \theta^\top \sigma(\cdot),
\end{equation*}
in which $\sigma:\R^n\to\R^\dth$ is a $\cC^1$ function with locally Lipschitz derivative. The selection criterion is then chosen as the weighted ``least-squares'' functional 
\begin{equation}\label{d:ls_J}
\cJ_{(\tau,w)} (j,\theta) := \sum_{i=0}^{j-1}\mu^{j-i-1}|\vep(\theta,x(t^i))|^2 + \theta^\top R \theta,
\end{equation}
obtained by letting in \eqref{d:J} $\costf(i,j,s):=\mu^{j-i-1}|s|^2$ and $\regf(\theta):=\theta^\top R\theta$, with $\mu\in[0,1)$ and $R\in\SPD_\dth$ design parameters called, respectively, the \emph{forgetting factor}  and the \emph{regularization matrix}, where we denoted by $\SPD_\dth$ the set of positive semi-definite symmetric matrices of dimension $\dth\x\dth$. Further remarks on $\mu$ ad $R$ are postponed at the end of the section.

The state-space of the identifier is the set $\cZ:=\SPD_\dth\x\R^\dth$. For a $z\in\cZ$ we consider  $z=(z_1,z_2)$, with $z_1\in\SPD_\dth$ and $z_2\in\R^\dth$, and equip $\cZ$ with the norm $|z|:=|z_1|+|z_2|$. We construct the identifier by relying on the following assumption.
\begin{assumption}\label{ass:PE}
There exist $\epsilon>0$ and, for every solution pair $((\tau,x),d)$ to \eqref{s:core} with $x(0)\in\cX_0$ and $d\in\D(x(0))$, a $j\sr\in\N$, such that for all $j\ge j\sr$ 
\begin{equation*} 
\msv\left(R+ \sum_{i=0}^{j-1}\mu^{j-i-1}\sigma(x(t^i))\sigma(x(t^i))^\top \right) \ge \epsilon  .
\end{equation*}
\end{assumption}

With   $X\sr$ and $\Xi\sr$   the compact sets defined in Section \ref{sec:extObs}, we then assume the following.
\begin{assumption}\label{ass:ls}
Assumption \ref{ass:x} holds with $\cX$ satisfying $\cX\subset X\sr$ and $\phi(\cX)\subset \Xi\sr$.
\end{assumption} 
Define
\begin{equation}\label{d:ls_cc}
\begin{aligned}
c_1 &:= (1-\mu)\inv \sup_{x\in X\sr}|\sigma(x)\sigma(x)^\top |\\
c_2 &:= (1-\mu)\inv \sup_{(x,y)\in X\sr\x \Xi\sr}|\sigma(x)y|,
\end{aligned}
\end{equation}
and, with $\epsilon$ given by Assumption \ref{ass:PE}, define the compact set
\begin{equation*} 
\rZ\sr  := \Big\{ z\in\cZ\st  \msv(z_1+R)\ge \epsilon,\, |z_1|\le c_1,\,|z_2|\le c_2 \Big\}. 
\end{equation*}

Finally,  with $\cdot\pinv$ denoting the Moore-Penrose pseudoinverse, let $\Sigma:\R^n\to\SPD_\dth$, $\lambda:\R^n\x\R\to\R^\dth$ and $\thmap:\cZ\to\R^\dth$ be  continuous functions satisfying\footnote{We stress that, as in the construction of $\psi$ in Section \ref{sec:extObs}, the maps $\Sigma$, $\lambda$ and $\gamma$ in \eqref{d:ls_Sig_lam_thm} are obtained by saturating the maps $\ain\mapsto \sigma(\ain)\sigma(\ain)^\top$, $(\ain,\aout)\mapsto \sigma(\ain)\aout$ and $z\mapsto (z_1+R)\pinv z_2$ respectively on the compact sets $X\sr$, $X\sr\x\Xi\sr$ and $\rZ\sr$. In turn, \eqref{d:ls_Sig_lam_thm} needs to hold only inside those compact sets, and \eqref{d:ls_bounds_Sig_lam_thm} is possible by continuity. }
\begin{equation}\label{d:ls_Sig_lam_thm}
\begin{array}{lcl}
\Sigma(\ain) &=& \sigma(\ain)\sigma(\ain)^\top\\
\lambda(\ain,\aout) &=& \sigma(\ain)\aout\\
\thmap(z) &=& (z_1+R)\pinv z_2
\end{array}
\end{equation}
respectively in the sets $X\sr$, $X\sr\x\Xi\sr$ and $\rZ\sr$,  and
\begin{align}\label{d:ls_bounds_Sig_lam_thm}
|\Sigma(\ain)|&\le c_\Sigma,& |\lambda(\ain,\aout)|&\le c_\lambda,& |\thmap(z)|&\le c_{\thmap}
\end{align}
everywhere else,  for some $c_\Sigma,c_\lambda,c_{\thmap}>0$.

Then, the identifier is  described by the equations 
\begin{equation}\label{s:ls_z}
\begin{array}{lcl}
z_1^+ &=& \mu z_1 + \Sigma(\ain) \\
z_2^+ &=& \mu z_2 + \lambda(\ain,\aout)\\
\theta &=&   \thmap(z)  ,
\end{array}
\end{equation} 
which correspond to take in \eqref{s.identifier}, $\vhi(z,\ain,\aout):= \big( \mu z_1+\Sigma(\ain),\, \mu z_2+\lambda(\ain,\aout)\big)$. Then, the following result (proved in Appendix \ref{sec:apd:proof_ls}) holds.
\begin{proposition}\label{prop:ls}
	Suppose that Assumptions \ref{ass:PE} and \ref{ass:ls} hold. 
	Then the identifier \eqref{s:ls_z} satisfies the Identifier Requirement relative to \eqref{d:ls_J}.
\end{proposition}

\begin{remark}	
We   remark that  Assumption~\ref{ass:PE} is always guaranteed, with $\epsilon= \msv(R)$ and $j\sr=0$ for every solution, whenever the regularization matrix $R$ is positive definite. Nevertheless, the regularization term introduces a bias on the parameter estimate, in the sense that, in the  case in which a true model exists in the model set $\cM$, the ``true parameter'' $\theta_{\rm T}$ is guaranteed to be a minimum of \eqref{d:ls_J} only if $\theta_{\rm T}\in\ker R$. This, in turn, implies that taking $R$ nonsingular makes the identifier \eqref{s:ls_z} to converge only to a neighborhood of $\theta_{\rm T}$ whose size is related to the eigenvalues of $R$ (and thus can be made arbitrarily small). Thus, a compromise in the choice of $R$ is necessary, and we refer to the pertinent literature for further details \cite{Sjoberg1993,Sjoberg1995,Ljung1999}. Moreover, we observe that, when $R=0$, Assumption~\ref{ass:PE}  boils down to a  \emph{persistence of excitation} condition on $x$.
\end{remark}
\begin{remark}\label{rmk:mu}
	The role of the forgetting factor $\mu$ can be interpreted in two (strongly related) ways. First, in view of~\eqref{d:ls_J}, taking~$\mu$ small means giving less importance to older errors in the sum, thus favoring those values of $\theta$ that better fit  newer samples. The more $\mu$ is close to $1$, instead, the more the importance of the past samples equalizes. Second,~$\mu$ defines the \emph{learning dynamics} of \eqref{s:ls_z}, i.e. how fast the state~$z$ converges and forgets the initial conditions. It is interesting to notice that faster dynamics (smaller $\mu$) are thus associated with learning of more ``volatile'' information (the recent samples weight more), while slower dynamics (larger $\mu$) are associated with learning of ``long term'' information drawn by weighting more equally the whole history of samples.
\end{remark}

We conclude this section with a simple academic example aimed at illustrating Remark  \ref{rmk.EIV}   in the case of least squares.
\begin{example}\label{ex.noisy_regressor}
In \eqref{s:xx}, let $n=1$ and $\phi(x)=\theta_{\rm T} x$, for some $\theta_{\rm T}\in\R$. Consider the least squares identifier \eqref{s:ls_z}   constructed   with $\dth=1$, $\sigma$ the identity, $R=0$, $\mu=1/2$, $X\sr=[-10,\,10]$, $\Xi\sr=[-50,50]$, and $\epsilon=1/10$, and consider the interconnection \eqref{s:core_identifier} of the identifier with the core system, in which the identifier is provided with the inputs $\ain(j) = x(t^j)+\din(j)$ and $\aout(j)=\phi(x(t^j))+\dout(j)$. 

First, notice that, in this setting, Assumption \ref{ass:PE} reduces to ask  that for each solution to \eqref{s:core_identifier} with $x(0)\in\cX_0$ and $d\in\D(x(0))$ there exists $j\sr\in\N$ such that, for all $j\ge j\sr$,
\begin{equation}\label{ex.ls.pe}
\sum_{i=0}^{j-1} \mu^{j-i-1} x(t^i)^2 \ge \epsilon .
\end{equation} 
Moreover, as clear from the proof of Proposition~\ref{prop:ls}, for every such solution  the corresponding steady-state trajectory~$z\sr$   satisfies  
$
z_1\sr (j) = \sum_{i=0}^{j-1} \mu^{j-i-1} x(t^i)^2
$ and $
z_2\sr (j) = \sum_{i=0}^{j-1} \mu^{j-i-1} x(t^i)\phi(x(t^i))=\theta_{\rm T}z_1\sr(j)
$.
Thus, \eqref{ex.ls.pe} is equivalent to 
\begin{equation}\label{ex.ls.z1_ge_eps}
z_1\sr(j) \ge \epsilon 
\end{equation}
for all  $j\ge j\sr$. Let $c_2>0$ be such that $|z_2\sr(j)|\le c_2$ for all $j\in\N$. Then, in view of \eqref{d:ls_Sig_lam_thm},  the condition \eqref{ex.ls.z1_ge_eps} implies
\begin{equation}\label{ex.ls.thsr_constr}
|\theta_{\rm T}| \le c_2/\epsilon  .
\end{equation}
 Therefore, Assumptions \ref{ass:PE} and \ref{ass:ls} impose  an implicit boundedness constraint  on the unknown parameter $\theta_{\rm T}$. It is worth noting, moreover, that Assumption~\ref{ass:PE} is used in the proof of Proposition \ref{prop:ls} to prove the regularity property of the Identifier Requirement. In turn, the constraint \eqref{ex.ls.thsr_constr} is  the manifestation of Remark \ref{rmk.EIV} into this specific context. 

Assume now   $\theta_{\rm T}=0$  and $x(j)=1$ for all $j\in\N$. Then, Assumption \ref{ass:PE} holds. Furthermore, suppose that the   disturbances $\din$ and $\dout$  satisfy $\din(j) = a v_{\rm in}(j)$ and $\dout(j)=av_{\rm out}(j)$, with $v_{\rm in},\,v_{\rm out}:\N\to[-1/2,1/2]$ and with  $a\in[0,1]$. Then,  $x(j)+\din(j)\in X\sr$ and $\phi(x(j))+\dout(j)\in\Xi\sr$ for all $j\in\N$. Thus, if $z(0)=0$, for $j\ge j\sr$ the parameter estimate satisfies
\begin{equation*}
\theta(j;a) = \dfrac{\sum_{i=0}^{j-1} \mu^{j-i-1} \big( 1 + av_{\rm in}(i) \big) \big( \theta_{\rm T} + a v_{\rm out}(i) \big) }{\sum_{i=0}^{j-1} \mu^{j-i-1} \big( 1 + av_{\rm in}(i) \big)^2},
\end{equation*}
and, hence, the presence of the disturbances $\din$ and $\dout$ in the variables induces a \emph{bias} in the parameter estimate. Nevertheless, we also notice that, for each fixed $j\ge j\sr$,  $\lim_{a\to 0} \theta(j;a)= \theta_{\rm T}$. Namely,  the bias vanishes continuously with the disturbances. Moreover, we can write
\begin{align*}
|\theta(j;a)-\theta_{\rm T}|&  =\Bigg|  \dfrac{\sum_{i=0}^{j-1} \mu^{j-i-1} \big(-av_{\rm in}(i)\theta_{\rm T} + av_{\rm out}(i)\big)}{\sum_{i=0}^{j-1} \mu^{j-i-1} \big( 1 + av_{\rm in}(i) \big)^2} \\&\!+ \dfrac{\sum_{i=0}^{j-1} \mu^{j-i-1} \big(a^2 v_{\rm in}(i) (v_{\rm out}(i)-v_{\rm in}(i)\theta_{\rm T}) \big) }{\sum_{i=0}^{j-1} \mu^{j-i-1} \big( 1 + av_{\rm in}(i) \big)^2}\Bigg|\\
&\le \kappa |a| \le \kappa \sup_{j\in\N}|(\din(j),\dout(j))|,
\end{align*}
with $\kappa=3(c_2/\epsilon+1)/(1-\mu)$, in which we used \eqref{ex.ls.thsr_constr}, and the fact that $\sum_{i=0}^{j-1}\mu^{j-i-1}(1+a v_{\rm in}(i))^2 \ge 1/4$. Therefore, according to the latter inequality, the presence of the bias is not in contrast with the Identifier Requirement. 
\end{example}

%
\subsection{Universal Approximators via Multiresolution Wavelet Identifiers and Cascade Least Squares}\label{sec:identifiers:wav}
In this section, we design an identifier   approximating arbitrary continuous $L^2$ functions with compact support. The identifier implements a ``chain'' of least-squares identifiers of the kind proposed in Section \ref{sec:identifiers:ls} performing a wavelet expansion of the inferred prediction model $\hat\phi$. The first least squares stage of the chain captures the best representation of the prediction model $\hat\phi$ at the coarsest scale; each successive least squares stage in the chain captures the information about the ``detail'' that is missing from the approximation attained by the previous stages to obtain an approximation of $\hat\phi$ at a finer scale.  We suppose the reader to be familiar with the wavelet theory. An essential review of the main tools used in this section is reported in Appendix \ref{sec:Wav}.  For convenience, we consider again the single variable case (i.e. $\dy=1$), and we recall  that a multivariable identifier can be obtained by the composition of multiple single variable identifiers.
To support the forthcoming construction, we assume\footnote{We remark that, in view of Assumption~\ref{ass:x}, there is no loss of generality in assuming $\phi\in\CC$ whenever the application of Section \ref{sec:observer} is concerned.} that the map $\phi$ of \eqref{s:xx} belongs to the set $\CC\subset L^2(\R^n)$ of compactly supported continuous functions in $L^2(\R^n)$.

As a first step, fix a GMRA $(\bV_i)_i$ associated to {\em compactly supported}  scaling function $\Scalf$ and wavelet functions $\Wavf^h$, $h=1,\dots, 2^{n}-1$, that are $\cC^1$ and with  Lipschitz derivative. Fix a \emph{starting scale} $i_0\in\Z$ and a \emph{target scale} $i_T\le i_0$ arbitrarily, and let $N:=1+|i_0-i_T|$. Let $\bH_{i_0}\subset\Z^n$ and $\bK_{i}\subset\Z^{n}$, $i=i_T+1,\dots,i_0$, be sets such that
\begin{equation*}
\begin{aligned} 
\supp\phi &\subset \bigcup_{\bh\in\bH_{i_0}} \supp \Scalf_{i_0,\bh}, &  \supp\phi  \subset& \bigcup_{\bk\in\bK_i}\bigcup_{h=1}^{2^n-1} \supp\Wavf_{i,\bk}^h
\end{aligned}
\end{equation*}
and, for all $\bh\notin \bH_{i_0}$, $\bk\notin\bK_{i}$ and $h=1,\dots,2^n-1$
\begin{equation*}
\begin{aligned} 
 \supp\phi  &\cap \Scalf_{i_0,\bh} =\emptyset, & \supp\phi  &\cap \Wavf_{i,\bk}^h =\emptyset .
\end{aligned}
\end{equation*}

The sets $\bH_{i_0}$ and $\bK_{i}$, $i=i_T+1,\dots,i_0$ exist and are finite due to the fact that $\phi$, $\Scalf$ and $\Wavf^h$ are in $\CC$. Moreover, they individuate $N$ \emph{finite} sets of functions, given by $\{\Scalf_{i_0,\bh}\st \bh\in\bH_{i_0} \}$ and $\{\Wavf_{i,\bk}^h\st h=1,\dots,2^n-1,\, \bk\in\bK_{i} \}$ for $i=i_T+1,\dots,i_0$, that can be used as regressors to construct the \emph{best} approximation of $\phi$ at scale $i_T$ as indicated in \eqref{wav:eq:wav_exp_Rd}. 

In particular, for $i=i_0$  let
\begin{equation}\label{d.wav.sig_i0}
\begin{array}{lcl} 
\sigma^{i_0}  &:=& \col\big( \Scalf_{i_0,\bh} \st \bh\in\bH_{i_0} \big) \\
\ddth^{i_0} &:=& \card{\bH_{i_0}},
\end{array} 
\end{equation}
and, for $i=i_T,\dots,i_0-1$,  define
\begin{equation}\label{d.wav.sig_i} 
\begin{array}{lcl}
\sigma^{i}   &:=&   \col\big( \Wavf_{i+1,\bk}^h \st h=1,\dots,2^{n}-1,\  \bk\in\bK_{i+1} \big)\\ 
\ddth^i &:=&  (2^{n}-1) \card{\bK_{i+1}}.
\end{array}
\end{equation}
For each $i=i_T,\dots,i_0$, pick $R^i\in\SPD_{\ddth^i}$ and $\mu^i\in[0,1)$ and assume the following\footnote{We stress that taking $R^i$ positive definite ensures that Assumption~\ref{ass:w:PE} holds with $j\sr=0$ for every solution.}.
\begin{assumption}\label{ass:w:PE}
For each $i=i_T,\dots,i_0$ there exist $\epsilon_i>0$ and, for every solution pair $((\tau,x),d)$ to \eqref{s:core} with $x(0)\in\cX_0$ and $d\in\D(x(0))$, a $j\sr_i\in\N$, such that for all $j\ge j\sr_i$ 
\begin{equation*}
\msv\left(R^i + \sum_{\ell=0}^{j-1} (\mu^i)^{j-\ell-1} \sigma^i\big(x(t^\ell)\big)\sigma^i\big(x(t^\ell)\big)^\top \right)\ge\epsilon_i.
\end{equation*}
\end{assumption}
Then, with $\cZ^{i}:=\R^{\ddth^{i}\x\ddth^{i}}\x\R^{\ddth^{i}}$,  define  for each $i=i_T,\dots, i_0$  a least square identifier $(z^i,\theta^i)\in\cZ^i\x\R^{\ddth^i}$ of the kind proposed in Section \ref{sec:identifiers:ls}  described by 
\begin{equation}\label{s:w_zi}
\begin{array}{lcl}
z_1^i{}^+ &=& \mu^i z^i_1 + \Sigma^i(\ain) \\
z_2^i{}^+ &=& \mu^i z^i_2 + \lambda^i(\ain,\eout^{i})\\
\theta^i{} &=&   \thmap^i(z^i)  ,
\end{array}
\end{equation} 
in which the quantities   $\Sigma^i,\ \lambda^i$ and $\thmap^i$ are constructed from~\eqref{d.wav.sig_i0}-\eqref{d.wav.sig_i} according to \eqref{d:ls_Sig_lam_thm}-\eqref{d:ls_bounds_Sig_lam_thm} by following, under Assumptions~\ref{ass:ls} and \ref{ass:w:PE}, the construction detailed in Section \ref{sec:identifiers:ls}, and the signal $\eout^i$ is defined as
\begin{equation}\label{d:w:eout}
\begin{array}{lcl}
\eout^{i_0} &:=& \aout\\
\eout^i &:=& \eout^{i+1} - (\theta^{i+1})^\top \sigma^{i+1}(\ain)\quad i=i_T,\dots,i_0-1.
\end{array}
\end{equation}

\begin{figure}[t]
	\centering
	\tikzstyle{sys} = [draw,inner sep=2,minimum width=3em,minimum height=2em]
	\tikzstyle{nosys} = [inner sep=2,minimum width=4em,minimum height=2.5em] 
	\tikzstyle{line} = [draw, -latex']
	\begin{tikzpicture}
	\node[sys] (s0) at (0,0) {$z^{i_0}$};
	
	\path[line] ($(s0.west)+(-2em,.5em)$)--($(s0.west)+(0,.5em)$) node[pos=0,right,anchor=east] {\scriptsize $\ain$};
	
	\path[line] ($(s0.west)+(-2em,-.5em)$)--($(s0.west)+(0,-.5em)$) node[pos=0,right,anchor=east] {\scriptsize $\aout$};

	%
	%

	\node[sys,right=2em of s0.east,yshift=1.5em] (s1)  {$z^{i_0-1}$};
	
	\path[line] (s0.east)-| ($(s1.west)+(-1em,-.5em)$) -- ($(s1.west)+(0,-.5em)$) node[pos=.5,below,yshift=-1em] {\scriptsize $\eout^{i_0-1}$};
	
	\path[line] ($(s0.west)+(-1em,.5em)$)|-($(s1.west)+(0,.5em)$);

	\path[line] (s1.east)--($(s1.east)+(1em,0)$) node[pos=1,below,xshift=1em] {\scriptsize $\eout^{i_0-2}$};
	
	\path[line] ($(s1.west)+(-1em,.5em)$)|-($(s1.east)+(1em,2em)$);

	\node[nosys,right=1.5em of s1.north east] (dots) {\large $\cdots$};

	\node[sys,right=2em of dots.north east] (sell)   {$z^{i_T}$};
	
	\path[line] ($(sell.west)+(-1em,.5em)$)--($(sell.west)+(0,.5em)$) node[pos=0,right,anchor=east] {\scriptsize $\ain$};
	
	\path[line] ($(sell.west)+(-1em,-1.5em)$)|-($(sell.west)+(0,-.5em)$) node[pos=0,below] {\scriptsize $\eout^{i_T}$};
	
	\path[line] (sell.east)--($(sell.east)+(1em,0)$);

	\end{tikzpicture}
	\caption{Chain structure of the scale-$i_T$ wavelet identifier.}
	\label{fig:wid}
	\vspace{-1em}
\end{figure}
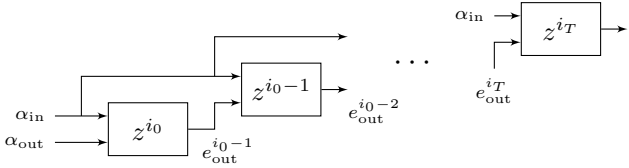

We then define the \emph{scale-$i_T$ wavelet identifier} as a system with state $z:=(z^{i_0},\dots,z^{i_T})\in\cZ:=\cZ^{i_0}\x\cdots\x\cZ^{i_T}$ and output $\theta:=\col(\theta^{i_0},\dots,\theta^{i_T})\in\R^\dth$, $\dth:=\ddth^{i_0}+\dots+\ddth^{i_T}$, given by the composition of \eqref{s:w_zi}-\eqref{d:w:eout} for $i=i_T,\dots,i_0$. We compactly rewrite it as 
\begin{equation}\label{s:w:wav_id}
\begin{array}{lcl}
z^+ &=& \vhi(z,\ain,\aout)\\
\theta &=&   \thmap (z)  ,
\end{array}
\end{equation} 
with $\vhi$ and $\thmap$ opportunely defined. Figure \ref{fig:wid} shows a block-diagram representation of the chain of least-squares identifiers composing \eqref{s:w:wav_id}.

The construction above is motivated by the following interpretation.
The first stage $z^{i_0}$ seeks the best (according to a criterion of the form \eqref{d:ls_J})   solution  to the inferred relation
\begin{equation*}
\phi  = P_{i_0}\phi = \sum_{\bh\in\bH_{i_0}} a_{i_0,\bh} \Scalf_{i_0,\bh},
\end{equation*} 
in the sense that its output $\theta^{i_0}$ will asymptotically give the optimal guess of the coefficients $a_{i_0,\bh}$, not known and ideally given by $\sprod{\phi}{\widetilde\Phi_{i_0,\bh}}$. Hence, the model $(\theta^{i_0})^\top\sigma^{i_0}$ represents the best approximation of $\phi$ at scale $i_0$, and the quantity
\begin{equation*}
\eout^{i_0-1} := \phi - (\theta^{i_0})^\top \sigma^{i_0}
\end{equation*}
represents the prediction error attained by the best scale-$i_0$ approximation of $\phi$. The second stage, $z^{i_0-1}$, takes as inputs $\ain$ and $\eout^{i_0-1}$ and it seeks the best solution to  
\begin{equation*}
\eout^{i_0-1} = \sum_{\bk\in\bK_{i_0}}\sum_{h=1}^{2^n-1} b_{i_0,\bk}^h \Wavf_{i_0,\bk}^h,
\end{equation*}
which corresponds to finding the best expansion of the scale-$i_0$ error $\eout^{i_0-1}$ in terms of the scale-$i_0$ wavelet functions $\Wavf_{i_0,\bk}^h$. In particular, the output $\theta^{i_0-1}$ will asymptotically give the best guess of the coefficients $b_{i_0,\bk}^h$, not known and ideally given by $\sprod{\phi}{\widetilde\Wavf_{i_0,\bk}^h}$. The resulting approximation of $\phi$, given by the function
\begin{equation*}
(\theta^{i_0})^\top \sigma^{i_0} + (\theta^{i_0-1})^\top \sigma^{i_0-1}
\end{equation*}
is then an approximation of $\phi$ at the finer scale $i_0-1$, and the corresponding prediction error
\begin{equation*}
\eout^{i_0-2} := \phi - (\theta^{i_0})^\top \sigma^{i_0}- (\theta^{i_0-1})^\top \sigma^{i_0-1}
\end{equation*}
represents the error attained by such best  approximation of $\phi$ at scale $i_0-1$, and provides the input to the next stage $z^{i_0-2}$. 

The same reasoning applies to all the successive scales up to the target scale $i_T$, whose  prediction error, expressed in the form \eqref{d:pred_error}, reads as
\begin{equation*}
\vep(\theta,x) = \phi(x) - (\theta^{i_0})^\top \sigma^{i_0}(x) -\  \ldots\  -(\theta^{i_T})\sigma^{i_T}(x)
\end{equation*}
and represents the prediction error attained by the best approximation of $\phi$ at scale $i_T$.

More formally, the identifier \eqref{s:w:wav_id} is characterized by the following lemma and the forthcoming proposition, proving that it satisfies the Identifier Requirement.
\begin{lemma} \label{lem:wav_pt1}
	Suppose that Assumptions~\ref{ass:ls} and \ref{ass:w:PE} hold. Then 
	there exist a compact set $\rZ\sr\subset\cZ$, $\beta_z\in\cK\cL$, $\rho_\theta,\rho_z>0$ and, for each solution pair $((\tau,x),d)$ to the core process \eqref{s:core}  with $x(0)\in\cX_0$ and $d\in\D(x(0))$, a hybrid arc $z\sr:\dom(\tau,x)\to\cZ$ and a $j\sr\in\N$, such that  $((\tau,x,z\sr),(d,\delta))$ with $\delta=0$ is a solution pair to \eqref{s:core_identifier} satisfying $z\sr(j)\in\rZ\sr$ for all $j\ge j\sr$, and the stability and regularity items of the Identifier Requirement hold. 
\end{lemma}
The proof of Lemma \ref{lem:wav_pt1} follows from Proposition \ref{prop:ls} by quite standard inductive arguments used to study cascade interconnections of stable systems and, for reasons of space, it is thus omitted. For $i=i_T,\dots,i_0$, let $z^i{}\sr$ and $\theta^i{}\sr$ be such that the hybrid arc  $z\sr$ produced by Lemma~\ref{lem:wav_pt1} and $\theta\sr:=\gamma(z\sr)$  satisfy  $z\sr=(z^{i_0}{}\sr,\dots,z^{i_T}{}\sr)$ and $\theta\sr=(\theta^{i_0}{}\sr,\dots,\theta^{i_T}{}\sr)$.
 Proposition~\ref{prop:ls} then implies that the first stage $z^{i_0}$ also satisfies the optimality item of the Identifier Requirement relative to
 \begin{equation}\label{q:wid:J0}
 \cJ_{(\tau,x)}^{i_0}(j,\theta^{i_0}):=\sum_{\ell=0}^{j-1}(\mu^{i_0})^{j-\ell-1} |\vep^{i_0}(\theta^{i_0},x(t^\ell))|^2 + (\theta^{i_0})^\top R^{i_0}\theta^{i_0}
 \end{equation} 
 where we let
 \begin{equation*}
\vep^{i_0}(\theta^{i_0},x) := \phi(x)- (\theta^{i_0})^\top \sigma^{i_0}(x).
 \end{equation*}
 For convenience, let $\theta^{[i]}:=(\theta^{i_0},\dots,\theta^i)$, and define
 \begin{equation*}
 \vep^{i}(\theta^{[i]},x) := \phi(x)- \sum_{k=i}^{i_0} (\theta^k)^\top \sigma^k(x).
 \end{equation*}
Then, define the following recursion
\begin{equation*}
\begin{aligned}
&\cJ_{(\tau,x)}^i(j,\theta^{[i]}) := \cJ_{(\tau,x)}^{i+1}(j,\theta^{[i+1]}) \\ & + \sum_{\ell=0}^{j-1} (\mu^{i})^{j-\ell-1} \big|\vep^{i}\big( (\theta^{i_0}{}\sr(t^\ell),\dots,\theta^{i+1}{}\sr(t^\ell),\theta^{i}), x(t^\ell)  \big)\big|^2 \\&\qquad+ (\theta^i)^\top R^i \theta^i
\end{aligned}
\end{equation*}
starting with \eqref{q:wid:J0} for $i=i_0$.
By inductive arguments, and in view of Lemma \ref{lem:wav_pt1}, it is thus possible to conclude the following.
 \begin{proposition}\label{prop:wav}
 	Under the assumptions of Lemma \ref{lem:wav_pt1},  \eqref{s:w:wav_id} satisfies also the optimality requirement relative to $\cJ^{i_T}_{(\tau,x)}$. Hence, it satisfies the Identifier Requirement relative to $\cJ^{i_T}_{(\tau,x)}$.
 \end{proposition}
\begin{remark}
	It is worth noticing that the same approximation attainable by the identifier \eqref{s:w:wav_id} at scale $i_T$ could be in principle obtained  by a single least squares stage by simply letting $i_0=i_T$ (i.e. by directly starting from a finer scale).  However, with a single least-squares stage, changing resolution would result in a completely new identification procedure, in which all the parameters have to be estimated again from scratch. The proposed cascade identifier \eqref{s:w:wav_id}, instead, permits to add or remove detail stages, thus increasing or decreasing resolution, without affecting the parameter estimates of the coarser stages. In addition it also allows to separate the ``learning dynamics'' of each stage, that are characterized by $\mu^i$. 
\end{remark}
\begin{remark}
	We also observe that there is a somewhat ``natural'' ordering in the choice of the forgetting factors $\mu^i$. In fact, coarser scales (i.e. higher $i$) are usually associated to rougher, yet fundamental, traits and  finer scales correspond instead to more ``volatile'' details. In view of Remark \ref{rmk:mu}, indeed, this intuition justifies choosing  $\mu^{i_0}>\mu^{i_0-1}>\cdots>\mu^{i_T}$ as  {\em learning} of fundamental rough traits, associated to long-term memory, is slower to acquire and forget than details, associated instead to a short-term memory and a faster learning dynamics.
\end{remark} 
 
%
%
%
\section{Examples}\label{sec:example} 
\subsection{Wavelet Identification}
We present here an academic example in which the adaptive observer developed in the previous sections is used to estimate the state and model of the following nonlinear oscillator
\begin{equation}\label{ex:s:x}
\dot x_1 =  x_2,\qquad \dot x_2 = \phi(x_1) ,
\end{equation}
by means of the wavelet identifier presented in Section \ref{sec:identifiers:wav}.

We suppose that Assumption~\ref{ass:x} holds with $\cX_0:=[-3,3]\x[-4,4]$ and $\cX:=\{x\in\R^2\st |x|\le 10\}$. This is the case, for instance, if $\phi$ is taken equal to one of the following two expressions
\begin{equation}\label{ex:phi12}
\phi_1(x_1) := 4 x_1 -  x_1^3,\quad\ \phi_2(x_1) :=3 \arctan(x_1) - x_1. 
\end{equation}
As a matter of fact, for each $k=1,2$,  the quantity
\begin{equation*}
H_k(x) := \dfrac{1}{2}x_2^2-\int_0^{x_1} \phi_k(s) ds  
\end{equation*}
is constant along each solution of \eqref{ex:s:x}
 obtained with $\phi=\phi_k$. As $H_k$ is continuous,  $H_k(\cX_0)$ is compact, and since
 \begin{equation*}
 \begin{aligned}
 \int_0^{x_1} \phi_1(s) ds &= 2x_1^2 -\dfrac{1}{4} x_1^4    \\
 \int_0^{x_1} \phi_2(s) ds &=  - \left( x_1^2/2+  \log(x_1^2+1)\right) + 2 x_1 \arctan(x_1),
 \end{aligned}
 \end{equation*} 
then in both cases  every solution to \eqref{ex:s:x} originating in $\cX_0$ satisfies  $H_k(x(t))\le \max_{x\in\cX_0,k=1,2}H_k(\cX_0) < 11$. Noticing that, by construction, $|x_1|\ge 10$ implies $|x|^2 \le 2 H_1(x)$ and  $|x|^2\le 4 H_2(x)$, then we can conclude that each solution to~\eqref{ex:s:x} originating in $\cX_0$, and obtained with either $\phi=\phi_1$ or $\phi=\phi_2$,   satisfies $|x(t)|\le \sqrt{\max\{2\cdot 11,4\cdot 8\}} < 6$, for all $t$, so that  $x(t)\in\cX$.

\begin{figure}[h!]
	\centering
	\includegraphics[width=\linewidth,trim=3.2em 0.3em 3.2em .1em,clip]{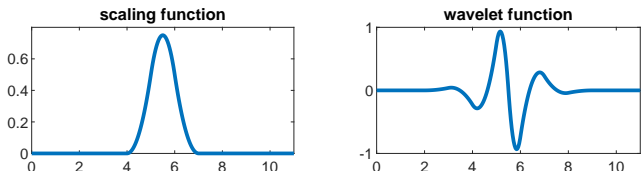}
	\caption{Scaling function and wavelet function.}
	\label{fig:scawav}\vspace{-.8em}
\end{figure}
\begin{figure}[h!]
	\includegraphics[width=\linewidth,clip,trim=3.2em 2.5em 3.2em 1em]{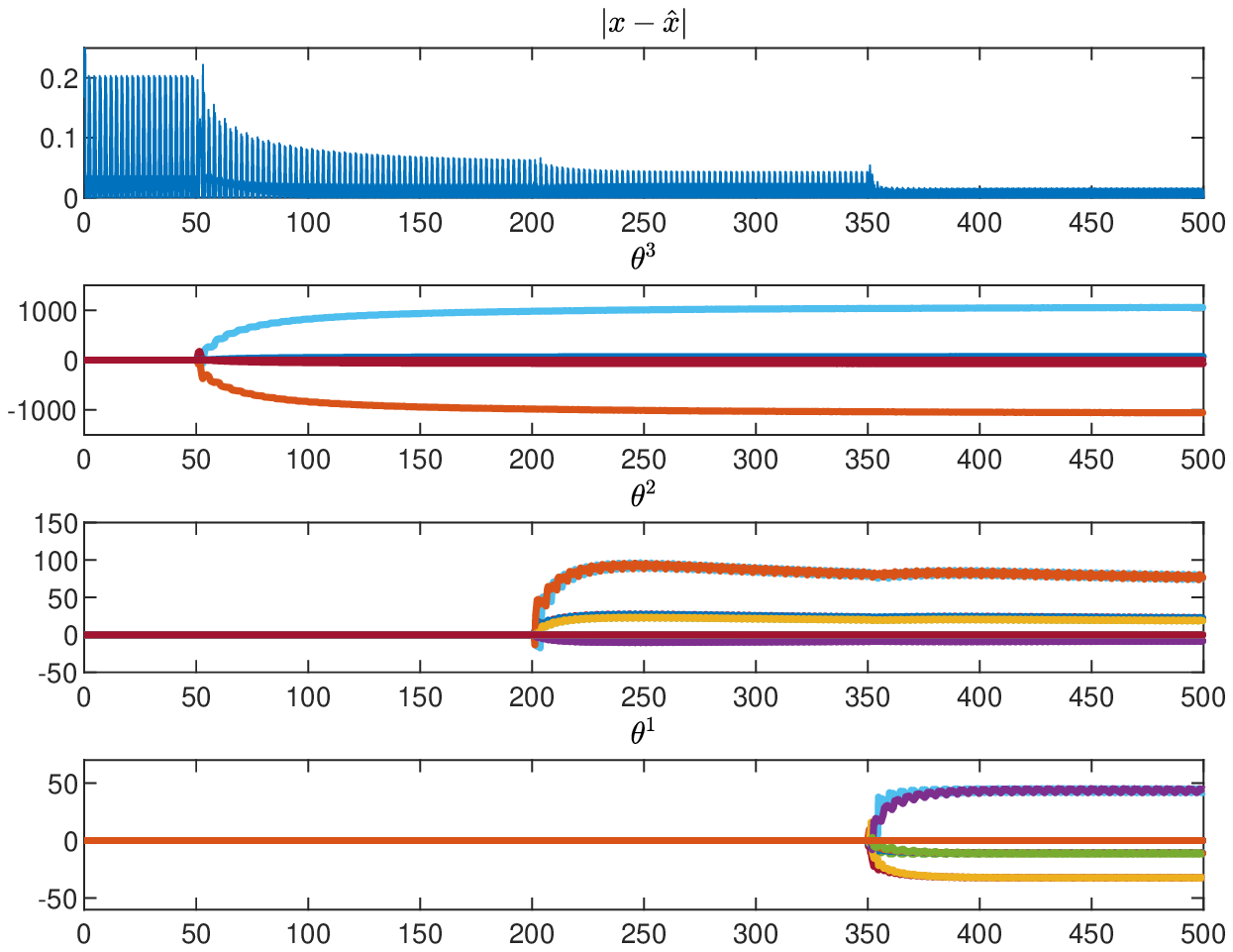}
	\caption{Time evolution of the state-estimation error $x-\hat x$ and of the identifier's parameters $\theta^i$ in the first case in which $\phi=\phi_1$.}\vspace{-.8em}
	\label{fig:est_err} 
\end{figure}
\begin{figure}[h!]
	\includegraphics[width=\linewidth,clip,trim=3.2em 2.5em 3.2em 1em]{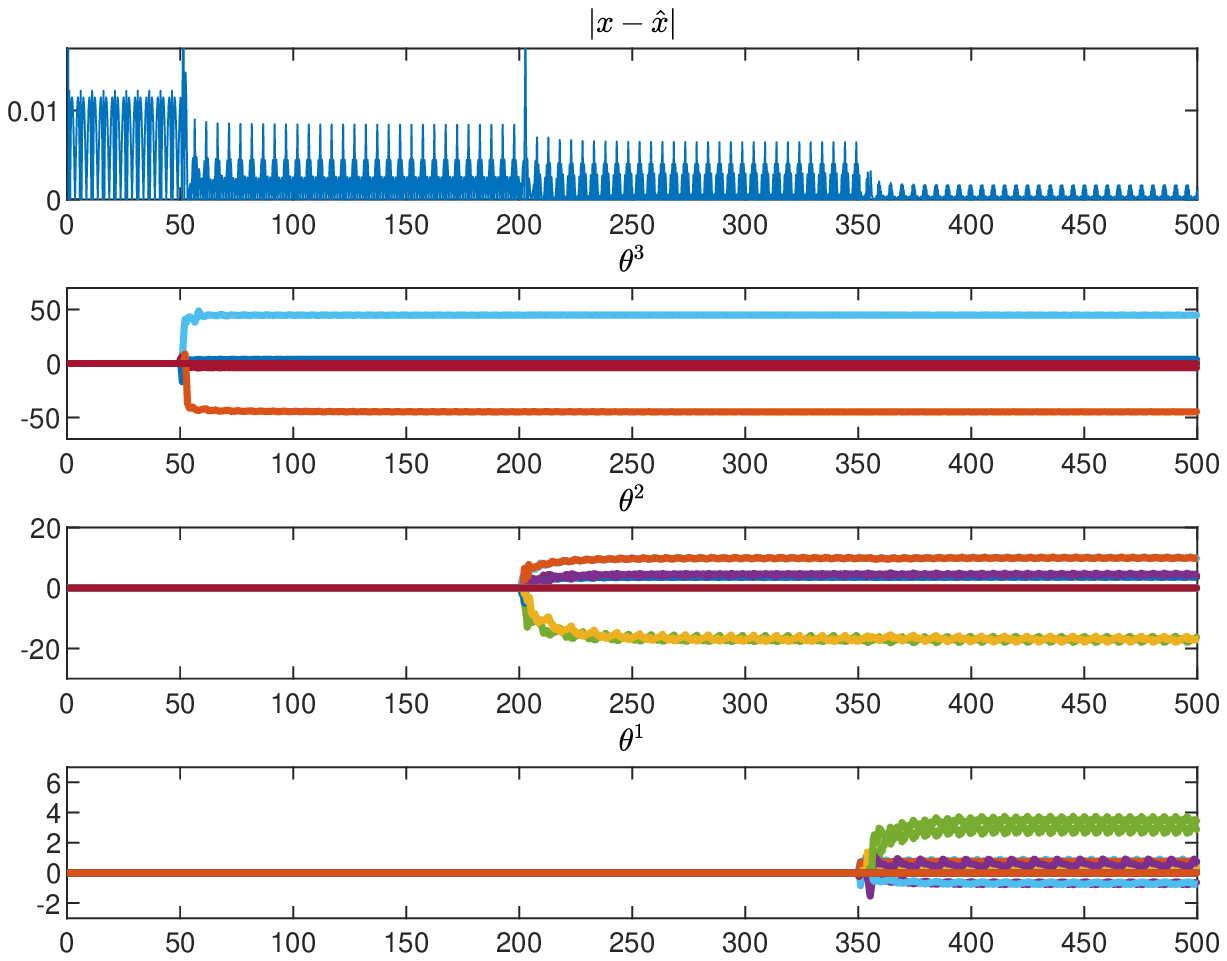}
	\caption{Time evolution of the state-estimation error $x-\hat x$ and of the identifier's parameters $\theta^i$ in the second case in which $\phi=\phi_2$.}
	\label{fig:est_err2} \vspace{-.8em}
\end{figure}
\begin{figure}[h!]
	\includegraphics[width=\linewidth,clip,trim=3.2em .2em 3.2em .0em]{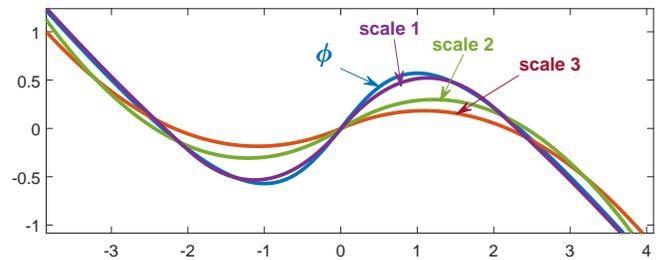}
	\caption{Approximation of $\phi=\phi_2$ at the different scales.}
	\label{fig:approx} \vspace{-.8em}
\end{figure}

The knowledge of $\cX$ and $\cX_0$, defined above, allows us to construct  an adaptive observer as specified in Section \ref{sec:observer} and Theorem \ref{thm:main}, by taking $X\sr$ so that $\cX\subset\rX\sr$ and $\Xi\sr$ large enough to handle all the functions $\phi$ of interest (for instance $\Xi\sr:=\{x\in\R^2\st|x|\le 10^3 \}$ allows to handle both~\eqref{ex:phi12}). 
For the identification phase, we use the multiresolution  identifier of Section \ref{sec:identifiers:wav} to perform a wavelet expansion of~$\phi$. Since~$\phi$ depends only on $x_1$, for ease of exposition we limit to {1-dimensional} wavelets, i.e. in   Section \ref{sec:identifiers:wav}  we consider functions $\hat\phi(\theta,\cdot)$ inside $L^2(\R)$ depending only on $x_1$. We choose a biorthogonal B-spline construction\footnote{They can be obtained in MATLAB with the command \texttt{wavefun} of the Wavelet Toolbox with argument `\texttt{bior3.5}' and then taking the dual functions.} for the scaling and wavelet functions, both represented in Figure \ref{fig:scawav}, and we fix $i_0:=3$ as the starting scale and $i_T:=1$ as the target one, requiring thus the use of $N=3$ least-squares stages.

In the presented simulations, the three stages are added   progressively at successive instants of time. For the first $50$ seconds no identifier is present, and a canonical high-gain observer obtained with $\psi=0$ is thus used.    At time $50s$, a first stage $z^{3}$ is added, working at the initial scale $i_0:=3$, and the function~$\psi$ is modified accordingly. At time $200$s, a second stage $z^{2}$ is added to obtain a representation at scale $i_0-1=2$. Finally, at time $350$s, a third stage is added to reach a representation at scale $i_T=1$. 
The identifier samples and updates every $T=0.1$ seconds. To ensure that Assumption~\ref{ass:w:PE} holds, we set the regularization matrices of each stage to $R^{i}=10^{-3}I_{\ddth^i}$. The forgetting factors are chosen as $\mu^{3}:=0.999$, $\mu^{2}:=0.995$, and $\mu^{1}:=0.99$. The design of the observer is then concluded by taking $g=25$.

Two simulations have been obtained by applying the same observer to \eqref{ex:s:x} with $x(0)=(-2.5,3)$ and with $\phi$ taken equal to $\phi_1$ and $\phi_2$ of \eqref{ex:phi12} respectively. Figures \ref{fig:est_err} and \ref{fig:est_err2} show  the time evolution, in both cases, of the norm of the estimation error $x(t)-\hat x(t)$ and of the output  $\theta^i(j)$ of each least-squares stage. Figure \ref{fig:approx} shows  instead the corresponding approximation of the function $\phi_2$  given by the  models at the different scales associated with the final value of the estimated parameters.



\subsection{Least-Squares Estimation With Measurement Noise}
According to the second claim of Theorem \ref{thm:main}, and as typical of high-gain designs, measurement noise is the disturbance having the worst effect on the estimates, as it gets amplified by a worst-case factor of $g^n$. Nevertheless, according to the claim of Theorem \ref{thm:main}, the measurement noise acts ``continuously'' on the asymptotic bound on the state and parameter estimates. Therefore we may expect such estimates to behave still reasonably well for small enough noise, where ``small enough'' depends on the fixed control parameter $g$.

By this example we aim to show how the measurement noise affects the performance of parameter and state estimation when the proposed adaptive observer is used together with the least-squares identifier of Section \ref{sec:identifiers:ls}. 

We consider the following class of systems
\begin{equation}\label{s.ex2}
\begin{array}{lcl}
\dot x_1 &=& x_2\\
\dot x_2 &=& x_3\\
\dot x_3 &=& \alpha x_2 + 3\beta x_1^2 x_2 + \ell( (1-x_1^2)x_3 -2 x_1 x_2^2) \\
y &=& x_1+\nu
\end{array}
\end{equation}
in which $\theta_{\rm T}:=(\alpha,\beta,\ell)\in\R^3$ is an unknown parameter, and~$\nu$ is a disturbance acting on the measured output $y$. For different initial conditions and values of $\theta_{\rm T}$, System~\eqref{s.ex2} models different  linear and nonlinear oscillators, as for instance the (extended) Duffing and Van der Pol oscillators \cite{Forte2013}.

We suppose that $\theta_{\rm T}$ ranges in a compact set  $\Theta\sr\subset[-5,\,5]\x[-5,0]\x[0,\,5]$ whose elements are such that Assumption~\ref{ass:x} holds with  $\cX_0=[-2,\,2]^3$ and $\cX:=[-10,\,10]\x[-10,\,10]\x[-100,\,100]$. 
Based on this, in the following simulations we have chosen the observer  degrees of freedom as follows:
\begin{enumerate}
	\item [(i)]  In \eqref{s:obs}-\eqref{d:Ct_Dt}, we let $\uT=\pT=1/2$,  and we chose $\Lambda_1$ and~$\Lambda_2$ according to \eqref{d.Lambdas}, with $K=(4,\,6,\,4,\,1)$.
	\item[(ii)] We  defined the identifier's data $(\cM,\cZ,\vhi,\thmap)$ as specified in the least squares section (Section \ref{sec:identifiers:ls}) with $\dth=3$, $\mu=0.995$, $R=0$, $\sigma=\col(\sigma_1,\,\sigma_2,\,\sigma_3)$,  with $\sigma_1(x):= x_2$, $\sigma_2(x)= 3x_1^2 x_2$, $\sigma_3(x):= (1-x_1^2)x_3-2x_1x_2^2$,
	and with $(\vhi,\gamma)$   chosen according to \eqref{s:ls_z} with
	\begin{align*}
	\Sigma(\ain)&:= \sat_{c_\Sigma}\big( \sigma(\ain)\sigma(\ain)^\top  \big) \\
	\lambda(\ain,\aout)&:= \sat_{c_\lambda}\big( \sigma(\ain)\aout  \big)\\
	\gamma(z)&:= \sat_{c_\gamma}\big(z_1\pinv z_2\big),
	\end{align*}
	in which $\sat$ denotes the component-wise saturation function\footnote{Namely, for a matrix $A\in\R^{n\x m}$ and a constant $M>0$, $\sat_M(A)$ is the matrix whose $(i,j)$ element is given by $M_{ij}:=\min\{\max\{A_{ij},-M\},M\}$, where $A_{ij}$ denotes the $(i,j)$-th entry of $A$.}, and   $c_\Sigma=10^7$, $c_\lambda=10^8$ and $c_\gamma=10$.
	
	\item[(iii)] According to Section \ref{sec:extObs}, and to the least-squares choice   $\hat\phi(\theta,\cdot)=\theta^\top\sigma(\cdot)$, $\psi$ is chosen as
	\begin{align*}
	\psi(\theta,\hat x,\xi)&=\sat_{M}\left(\theta^\top \dfrac{\partial \sigma(\hat x)}{\partial \hat x} \col(\hat x_2,\hat x_3,\xi)\right) \\
	&=\sat_{M}\Big(\theta_1 \hat x_3+\theta_2\big(6\hat x_1\hat x_2^2 + 3\hat x_1^2\hat x_3\big) \\&\qquad\quad + \ell\big( (1-\hat x_1^2)\xi -6\hat x_1\hat x_2\hat x_3-2\hat x_2^3\big) \Big),
	\end{align*}
	with $M=10^3$.
	
	\item[(iv)] The design is concluded by choosing $g=25$.
\end{enumerate}

Taking $R=0$ permits to have an asymptotic state and parameter estimate when the noise is not present. However, as remarked in Section  \ref{sec:identifiers:ls}, it implies that Claim~2 of Theorem \ref{thm:main} applies only to the solutions of \eqref{s.ex2} carrying enough excitation. While $c_\gamma$ is chosen on the basis of the knowledge of $\Theta\sr$, the saturation constants $M$, $c_\Sigma$ and $c_\lambda$ are chosen large enough so that, with $\phi(x):=\theta_{\rm T}^\top\sigma(x)$, we have $\psi(\theta_{\rm T},x,\phi(x))= (\partial \phi(x)/\partial x)\col(x_2,x_3,\phi(x))$,  $\Sigma(x)=\sigma(x)\sigma(x)^\top$, and $\lambda(x,\phi(x))=\sigma(x)\phi(x)$  along the solutions to \eqref{s.ex2} of interest.

The following simulations have been obtained with $x(0)=(1,-1,0)$, $\tau(0)=0$, $\hat x(0)=(0,0,0)$, $\xi(0)=0$, $z_1(0)=I$, and $z_2(0)=0$. The unknown parameter $\theta_{\rm T}$ is initially set to
\begin{equation*}
\theta_{\rm T} = (\alpha_1,\beta_1,\ell_1) =(-1,0,1/2),
\end{equation*}
and later, around time $t=1000$, changed to
\begin{equation*}
\theta_{\rm T} = (\alpha_2,\beta_2,\ell_2) =(1,-1/2,0).
\end{equation*}

\begin{figure}
	\includegraphics[width=\linewidth,clip,trim=3.4em 1.5em 3.0em .5em]{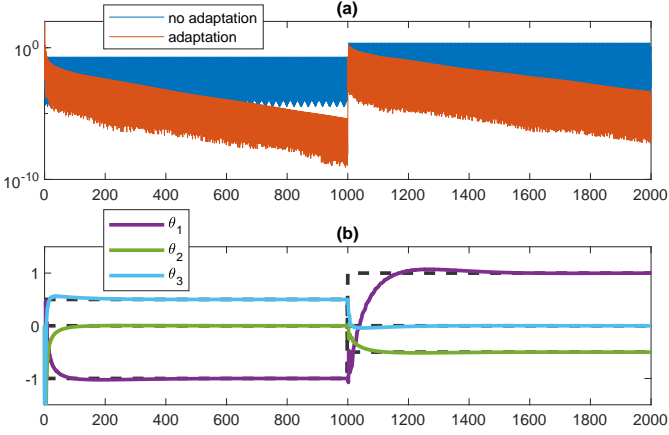}
	\caption{Comparison, in a noise-free setting, between the proposed adaptive observer and a non-adaptive extended high-gain observer obtained with $\psi=0$. Plot (a) shows, in semi-logarithmic scale, the time evolution of the norm $|\hat x-x|$ of the state estimation error. Plot (b) shows the time-evolution of the parameter estimate $\theta$ obtained by the adaptive observer. The values of $\theta_{\rm T}$ is depicted in dashed, gray line.}
	\label{fig:ex2_no_noise} \vspace{-.8em}
\end{figure}

In a first simulation, we compare,  in a noise-free setting, the   adaptive observer constructed above with a non-adaptive extended high-gain observer obtained with $\psi=0$. As shown in Figure~\ref{fig:ex2_no_noise},  with the adaptive observer the norm of the state estimation error $\hat x-x$ decays exponentially to zero (with a time constant determined by the parameter convergence), and the parameter~$\theta$ converges to $\theta_{\rm T}$. The non-adaptive observer instead shows a persistent estimation error.

\begin{figure}
	\includegraphics[width=\linewidth,clip,trim=4.5em 2.8em 2.5em 1.4em]{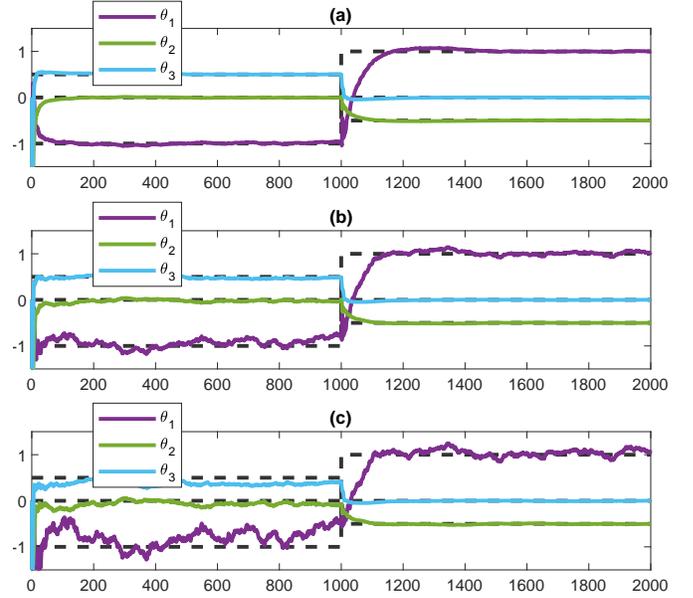}
	\caption{Time evolution of the parameter estimate with different noise levels. Plot (a) is obtained with $q=10^{-3}$, plot (b) with $q=5\cdot 10^{-3}$, and (c) with $q=10^{-2}$.}
	\label{fig:ex2_noise_params} \vspace{-.8em}
\end{figure}

In the next simulations we let $\nu$ be given by
\begin{equation*}
\nu(t) = q \nu_0(t),
\end{equation*}
in which $q>0$, and $\nu_0$ is obtained as a linear interpolation (in time) of the samples of an uniformly distributed stochastic process sampled every $0.1$ seconds, and taking values in $[-1/2,1/2]$.
Figure \ref{fig:ex2_noise_params} shows  the parameter estimates obtained with $q=10^{-3}$, $q=5\cdot 10^{-3}$ and $q=10^{-2}$ respectively. 
Clearly, as the noise amplitude increases, the parameter estimation performance decreases. Nevertheless, the continuity between  noise amplitude and  asymptotic estimation error claimed in Theorem \ref{thm:main} is evident from the plots.

\begin{figure}
	\includegraphics[width=\linewidth,clip,trim=4.5em 2.4em 2.5em 1.4em]{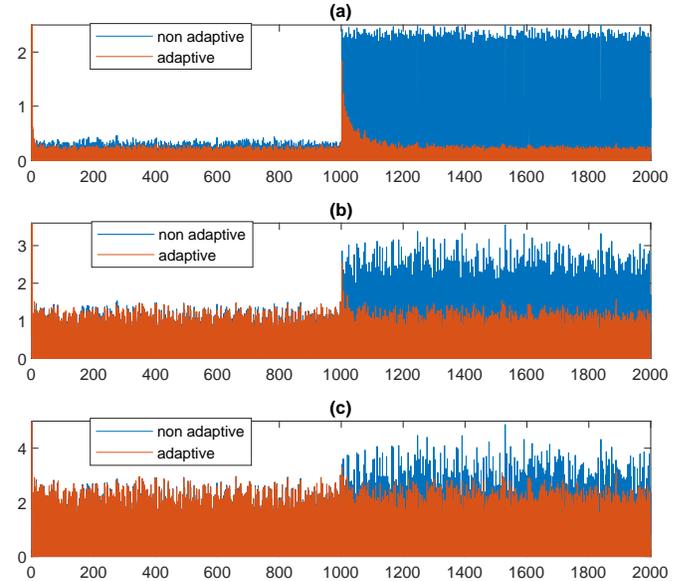}
	\caption{Time evolution of the state estimate $|\hat x-x|$ with different noise levels. Plot (a) is obtained with $q=10^{-3}$, plot (b) with $q=5\cdot 10^{-3}$, and (c) with $q=10^{-2}$.}
	\label{fig:ex2_noise_state} \vspace{-.8em}
\end{figure}

Finally, Figure \ref{fig:ex2_noise_state} shows the time evolution of the norm $|\hat x-x|$ of the state estimation error for the different noise levels. In presence of sufficiently small measurement noise,  the proposed adaptive observer  still performs considerably better than the non-adaptive one. However, as the noise amplitude increases, its effect on the state estimate becomes more important, dominating the effect of adaptation. Thus, as the noise amplitude increases, the performance of the adaptive observer gets close to the non-adaptive case, as evident in Figure \ref{fig:ex2_noise_state}.


%
%
\section{Conclusions}
In this paper we  considered the problem of joint state and model estimation for a class of nonlinear systems. We proposed an adaptive observer in which adaptation is cast and solved as a system identification problem. 
Practical  and approximate   estimation results are given, and robustness with respect to exogenous disturbances is provided in terms of an input-to-state stability property. Differently from canonical adaptive observer designs, here we did not assume a particular structure of the uncertainty and we did not propose an ad hoc adaptation mechanism. Rather,    different parametric and non-parametric system identification techniques can be applied. 
 Future research will focus on the extension of the framework to more general classes of systems and observers and to the application to output-feedback stabilization and regulation.

\appendix
\subsection{Proof of Theorem \ref{thm:main}}\label{apd:proofThm}
To prove the first claim, change coordinates as
\begin{equation*}
\hat x_i\mapsto \tilde x_i := g^{1-i}(\hat x_i-x_i)
\end{equation*}
and define $\zeta:=\col(\tilde x,g^{-n}\xi)$. Then $\zeta$ flows according to
\begin{equation*}
\dot\zeta =  g M \zeta + H_1(g)\psi(\theta,\hat x,\xi)  - H_2(g)d - H_3(g) \phi(x) + g\Lambda(1)\nu ,
\end{equation*}
in which  $\Lambda :=\col(\Lambda_1,\Lambda_2)$,    $H_1(g):=g^{-n}\col(0_\dy,B)$, $H_2(g):= \col\big(\diag(I_\dy,g\inv I_\dy,\dots, g^{1-n} I_\dy),\, 0_{\dy\x\dx}\big)$, and $H_3(g):=g^{1-n}\col(B,0_\dy)$, and $M$ is defined in \eqref{d:M}. Moreover, $\zeta$ jumps as $\zeta^+=\zeta$. Hence, in view of \eqref{bound:psi_omega}, and since by Assumption~\ref{ass:x} $\phi(\cX)$ is bounded and $x(t)\in\cX$ whenever $x(0)\in\cX_0$ and $d\in\D(x(0))$, standard high-gain arguments (see e.g. \cite{KhalilNL}) show that for some $a_0,a_1,a_2>0$, and  with $\dg$ defined in \eqref{d.oe_ee_dg},
\begin{equation*}
|\zeta(t,j)|\le  \max\Big\{   a_0  \e^{-a_1g t} |\zeta(0)| ,\,   g^{-n}a_0  ,\,   a_0|(\dg,\nu)|_{t}\Big\} 
\end{equation*}
holds for each $(t,j)\in\dom\zeta$,
from which the first claim of the theorem can be easily deduced.

Regarding the second claim, first  notice that, by definition of $\rC_\tau$ and $\rD_\tau$ in \eqref{d:Ct_Dt}, every  complete solution $\xb$ of \eqref{s:x_obs} has infinite jumps and flows, i.e. $\sup\jumps\xb=\sup\flows\xb=\infty$. Pick a complete solution pair $(\xb,(d,\nu))$ to \eqref{s:x_obs} and let $z\sr$ and~$\theta\sr$ be the corresponding hybrid arcs produced by the Identifier Requirement. Change  variables as
\begin{equation*}
\xi \mapsto \tilde\xi := g^{-n}(\xi - \hat\phi(\theta\sr,x)),
\end{equation*}
and let $\tilde \zeta:=\col(\tilde x,\tilde\xi)$. Then, noting that $\xi-\phi(x) = g^n\tilde \xi - \vep\sr$, it is readily seen that $\tilde\zeta$ flows according to
\begin{equation*}
\dot{\tilde\zeta} =  g M \tilde\zeta + H_1(g)\tilde\psi(\theta,\theta\sr,\tilde x,\tilde\xi,x)   - H_2(g)d - H_3(g) \vep\sr + g\Lambda(1)\nu ,
\end{equation*}
in which, by letting $G(g):=\diag(I_\dy,gI_\dy,\dots,g^{n-1}I_\dy)$,
\begin{align*}
 \tilde\psi(\theta,\theta\sr,\tilde x,\tilde\xi,x) &:= \psi\big(\theta,x+G(g)\tilde x, g^n\tilde\xi + \hat\phi(\theta\sr,x) \big) \\&\qquad- \dfrac{\partial \hat\phi(\theta\sr,x)}{\partial x}\big(Ax+B\phi(x)+ d\big).
\end{align*}
Furthermore, $\tilde \zeta$ jumps according to
\begin{equation*}
\tilde\zeta^+ = \tilde\zeta+  H_1(g)\big( \vep\sr{}^+-\vep\sr \big).
\end{equation*} 
 By the Identifier Requirement, there exists $j\sr\in\N$ such that $z\sr(j)\in\rZ\sr$ (and hence $\theta\sr(j)\in\thmap(\rZ\sr)\subset\Theta\sr$) for all $j\ge j\sr$. Thus, if $(x,\phi(x))$ is eventually in $X\sr\x\Xi\sr$,  there exists $\bar s\in\dom\xb$ of the form $\bar s =(t_{\bar j},\bar j)$ with $\bar j\in\jumps\xb$ satisfying $\bar j\ge j\sr$, such that $(x(t,j),\phi(x(t,j)),\theta\sr(t,j))\in X\sr\x\Xi\sr\x\Theta\sr$ for all $(t,j)\succeq\bar s$. Hence, in view of \eqref{d:psi_omega} and of the regularity item of the Identifier Requirement there exist $a_4,a_5>0$ such that 
\begin{equation*}
|\tilde\psi(\theta,\theta\sr,\tilde x,\tilde\xi,x)|\le a_4\big(  g^{n}|\tilde\zeta|  + |z-z\sr| + |\vep\sr|+ |d|  \big)
\end{equation*}
for all $(t,j)\succeq\bar s$. We thus conclude that there exist {$b_0,b_1,b_2>0$} and $g\sr_a >0$ such that, if $g\ge g\sr_a$, $\tilde\zeta$ satisfies
\begin{equation}\label{pf:bound_tildezeta}
\begin{aligned}
|\tilde \zeta(t,j)|&\le b_0 \max\Big\{ \e^{-b_1 g(t-t_j)}|\tilde\zeta(t_j,j)|,\, g^{-n}|\vep\sr|_{(t,j)},\\&\qquad\qquad\qquad g^{-(n+1)}|z-z\sr|_j,\, |(\dg,\nu)|_t \Big\}\\
|\tilde \zeta(t,j+1)|&\le  2\max\big\{ |\tilde\zeta(t,j)|,\, 2g^{-n} |(\vep\sr(t,j),\vep\sr{}(t,j+1))|\big\}  
\end{aligned}
\end{equation}
respectively during the flow and jump times succeeding $\bar s$, and with $\tilde\zeta(\bar s)$ that is bounded according to the first claim of the theorem.
We then observe that, since  for each $j\in\jumps\xb$  it holds that $t^j-t_j\ge \uT$, then for each constant $\bar b>0$, taking $g$ sufficiently large yields
\[
\lim_{t+j\to\infty} \bar b^{j+1} \e^{- b_1 g  t  } \le \lim_{j\to\infty} = \bar b^{j+1}\e^{-jb_1g\uT } =0.
\]
Therefore, there follows from \eqref{pf:bound_tildezeta} by induction and standard ISS arguments (see e.g. \cite{IsidoriNCS2}) that there exist $c_1,c_2>0$ and $g\sr_b(\uT)> g_a\sr$ such that $g\ge g_b\sr(\uT)$ implies
\begin{equation}\label{pf:bound_ls_zeta}
\begin{aligned}
\limsup_{t+j\to\infty}|\tilde\zeta| \le  \max\Big\{& c_1g^{-n} \limsup_{t+j\to\infty} |\vep\sr| ,\, c_2\limsup_{t\to\infty}|(\dg,\nu)| \\&\ c_2 g^{-(n+1)} \limsup_{j\to\infty} |z -z\sr | \Big\},
\end{aligned}
\end{equation}
where we used the fact that $\limsup_{t+j\to\infty}|\vep\sr|=\limsup_{t+j\to\infty}|\vep\sr{}^+|$.
On the other hand, the identifier is subject to the inputs $(\ain,\aout)=(\ain\sr+\din,\aout\sr+\dout)$ with $(\ain\sr,\aout\sr)=(x,\phi(x))$ and
\begin{align*}
\din &= G(g)\tilde x, & \dout&= g^n\tilde \xi - \vep\sr,
\end{align*}
in which we recall that  $G(g):=\diag(I_\dy,gI_\dy,\dots,g^{n-1}I_\dy)$.
Therefore, the stability item of the Identifier Requirement yields the estimate
\begin{equation}\label{pf:bound_ls_z}
\limsup_{j\to\infty}|z-z\sr| \le c_3\max\Big\{ g^n\limsup_{t+j\to\infty}|\tilde\zeta|,\, \limsup_{t+j\to\infty}|\vep\sr|  \Big\}
\end{equation}
with $c_3>0$   proportional to the Lipschitz constant of $\rho_z$. Pick 
\begin{equation*}
g>g\sr(\uT) := \max\big\{ g_b\sr(\uT),\,  c_1 ,\, c_3c_2  \big\},
\end{equation*}
then substituting \eqref{pf:bound_ls_z} into \eqref{pf:bound_ls_zeta}, and assuming without loss of generality $c_1\ge 1$, shows that 
\begin{equation}\label{pf:bound_tz_2}
\limsup_{t+j\to\infty}|\tilde\zeta| \le \max\Big\{ g^{1-n}\limsup_{t+j\to\infty}|\vep\sr|,\, c_2\limsup_{t\to\infty}|(\dg,\nu)|  \Big\},
\end{equation}
and substituting \eqref{pf:bound_ls_zeta} into \eqref{pf:bound_ls_z} yields
\begin{equation}\label{pf:bound_z_2}
\limsup_{j\to\infty}|z-z\sr|\le \max\Big\{\!c_4 \limsup_{t+j\to\infty}|\vep\sr|,  c_5g^n\limsup_{t\to\infty}|(\dg,\nu)| \!\Big\}
\end{equation}
with $c_4:=c_1c_3$ and $c_5:=c_2c_3$. Thus, the first equation of the second claim follows directly by \eqref{pf:bound_tz_2} and \eqref{pf:bound_z_2} in view of the regularity item of the Identifier Requirement. The second equation instead follows by \eqref{pf:bound_tz_2} once noted  that $|\hat x_i-x_i|\le g^{i-1}|\tilde\zeta|$. \hfill\IEEEQEDhere

\subsection{Proof of Proposition \ref{prop:ls}}\label{sec:apd:proof_ls}
By definition of $\Sigma$ and $\lambda$ in  \eqref{d:ls_Sig_lam_thm}, they are bounded and Lipschitz on $X\sr$ and $X\sr\x\Xi\sr$ respectively. Hence, there exists $\ell>0$ such that, for each $x_1\in X\sr$, $x_2\in\R^n$, $y_1\in\Xi\sr$ and $y_2\in\R$  it holds that 
\begin{equation}\label{pf:bound_Sig}
\begin{aligned}
|\Sigma(x_1)-\Sigma(x_2)|&\le \ell|x_1-x_2|, \\ |\lambda(x_1,y_1)-\lambda(x_2,y_2)|&\le \ell |(x_1,y_1)-(x_2,y_2)|.
\end{aligned}
\end{equation}

Pick a solution pair $((\tau,x),d)$ to \eqref{s:core} with $x(0)\in\cX_0$ and $d\in\D(x(0))$. Define  $z\sr=(z_1\sr,z_2\sr)$   by letting $z\sr(0)=(0,0)$ and, for $j\ge 1$,
\begin{align*}
z_1^\star(j) &:= \sum_{i=0}^{j-1}\mu^{j-i-1}\sigma(x(t^i))\sigma(x(t^i))^\top\\
z_2\sr(j) &:=\sum_{i=0}^{j-1}\mu^{j-i-1}\sigma(x(t^i))\phi(x(t^i)) .
\end{align*} 
Since by Assumption~\ref{ass:ls} $(x(t),\phi(x(t)))\in X\sr\x\Xi\sr$  then,  for $\delta=0$, $((\tau,x,z\sr),(d,\delta))$ is a solution pair to \eqref{s:core_identifier}. 

By construction, $|z_1\sr(j)|\le c_1$ and $|z_2\sr(j)|\le c_2$, with $c_1$ and~$c_2$ defined   in \eqref{d:ls_cc}.
By Assumption~\ref{ass:PE},  for $j\ge j\sr$, $\msv(z_1\sr(j)+R)\ge \epsilon$, so that $z\sr(j)\in \rZ\sr$ for all $j\ge j\sr$. 

Next, let   $\cJ_{(\tau,x)}$ given by \eqref{d:ls_J}. Then, for all $j\in\N$
\begin{align*}
\optmap_{(\tau,x)}(j) &= \left\{ \theta\in\R^\dth\st \dfrac{\partial \cJ_{(\tau,x)}(j,\theta)}{\partial \theta}=0 \right\}\\&=\Big\{ \theta\in\R^\dth\st (z_1\sr(j)+R)\theta=z_2\sr(j)\Big\}.
\end{align*}
As $z_2\sr(j)\in\img z_1\sr(j)$, and since   $z\sr(j)\in \rZ\sr$ implies    $\gamma(z\sr(j))=(z_1\sr(j)+R)\pinv z_2\sr(j)$, then $\theta\sr:=\gamma(z\sr)$ satisfies   $\theta\sr(j)\in\optmap_{(\tau,x)}(j)$ for all $j\ge j\sr$. Thus, the optimality requirement holds.

 Moreover, pick  a solution pair to \eqref{s:core_identifier} of the form $((\tau,x,z),(d,\delta))$, with $((\tau,x),d)$ the same as above. By direct solution we also obtain
\begin{align*}
&|z(j)-z\sr(j)|\le \mu^j |z(0)-z\sr(0)|\\& + \sum_{i=0}^{j-1}\mu^{j-i-1} \big( |\Sigma(x(t^i))-\Sigma(x(t^i)+\din(t^i))|\\& + |\lambda(x(t^i),\phi(x(t^i)))-\lambda(x(t^i)+\din(t^i),\phi(x(t^i))+\dout(t^i))|\big) 
\end{align*}
which in view of \eqref{pf:bound_Sig} yields
\begin{equation*}
|z(j)-z\sr(j)|\le 2\max\Big\{ \mu^j|z(0)-z\sr(0)|, 2\ell(1-\mu)\inv |\delta|_j \Big\}
\end{equation*}
which is the stability requirement.

Finally, since $\thmap$ is Lipschitz on $\rZ\sr$ and bounded, there exists $\rho_\theta>0$ such that, for all $(z,z\sr)\in \cZ\x\rZ\sr$, $|\thmap(z)-\thmap(z\sr)|\le \rho_\theta|z-z\sr|$, which implies the first part of the regularity item. The second part, instead, holds by definition of~$\sigma$.\null\hfill \IEEEQEDhere

%
%

\subsection{Basics of Wavelet Analysis}\label{sec:Wav}
For the basic concepts about wavelets, Riesz bases and biorthogonality, we refer to \cite{Daubechies1992,Walnut2002,Strang1996,Christensen2008}.
 A relevant framework in which biorthogonal wavelets can be constructed is the {\em Generalized Multiresolution Analysis (GMRA)} \cite{Walnut2002}.
\begin{definition}\label{def:GMRA}
	A GMRA on $\R$ is a sequence of subspaces $(V_i)_{i\in\Z}$ of $L^2(\R)$ satisfying the following properties:\footnote{Some authors (e.g. \cite{Walnut2002}) use higher values of $i$ to denote higher resolutions (i.e. $V_i\subset V_{i+1}$), some other (e.g. \cite{Daubechies1992}) use the opposite. We chose this latter convention according to \cite{Daubechies1992}.}
	\begin{enumerate}
		\item [a)] For all $i\in\Z$, $V_i\subset V_{i-1}$.
		\item [b)] For any $f\in\CCR$ and every $\epsilon>0$, there exist $i\in\Z$ and a function $g\in V_i$ such that $|f-g|\le \epsilon$. 
		\item [c)] $\cap_{i\in\Z} V_i = \{0\}$.
		\item [d)] $f\in V_i$ if and only if $f(2^i \cdot)\in V_0$.
		\item [e)] There exists a function $\scalf\in L^2(\R)$, called the {\em scaling function}, such that $V_0 = \cspan\{\scalf(\cdot-k)\}_{k\in\Z}$ and $\{\scalf(\cdot-k)\}_{k\in\Z}$ is a {\em Riesz basis} for $V_0$.
	\end{enumerate}
\end{definition}
For a function $f\in L^2(\R)$ and with $i,k\in\Z$, we define the function $f_{i,k}\in L^2(\R)$ as $f_{i,k}(s) = 2^{-i/2}f(2^{-i}s-k)$. If $\scalf$ is a scaling function of a GMRA $(V_i)_{i\in\Z}$, then  $V_i = \cspan\{\scalf_{i,k}\}_{k\in\Z}$ 
and $\{\scalf_{i,k}\}_{k\in\Z}$ is a Riesz basis for $V_i$. As the family $\{\scalf(\cdot-k)\}_{k\in\Z}$ is a Riesz basis of $V_0$, it admits a {\em dual basis} $\{\widetilde\scalf(\cdot-k)\}_{k\in\Z}$. If $\widetilde\scalf$ is the scaling function of a GMRA $(\widetilde V_i)_{i\in\Z}$ we say that the GMRAs $(V_i)_{i\in\Z}$ and $(\widetilde V_i)_{i\in\Z}$ are {\em dual} to each other. With $i\in\Z$ and $f\in L^2(\R)$ we define the {\em projection} operator $\PP_i$ and the {\em detail} operator $\QQ_i$ as
\begin{equation*}
\PP_i f  :=\sum_{k\in\Z} \sprod{f}{\widetilde\scalf_{i,k}}\scalf_{i,k},\qquad \QQ_if :=\PP_{i-1}f-\PP_if .
\end{equation*}
For a given $i\in\Z$, $\QQ_{i}f$ represents the additional detail that is needed to obtain a \textit{finer} approximation $\PP_{i-1}f$ of $f$ at scale $i-1$ starting from a \textit{coarser} approximation $\PP_{i}f$ at scale $i$. Moreover, for every $f\in \CCR$,  $\lim_{i\to-\infty}|\PP_if-f|=0$.
%

We can univocally associate with the scaling functions $\scalf$ and $\widetilde \scalf$, respectively, two functions $\psi$ and $\widetilde\psi$ in $L^2(\R)$, called the {\em wavelet functions}. With $W_i:=\cspan\{\wavf_{i,k}\}_{k\in\Z}$ and $\widetilde W_i:=\cspan\{\wavf_{i,k}\}_{k\in\Z}$, they fulfill the following properties.
\begin{enumerate}
	\item[a)] $\wavf\in V_{-1}$ and $\widetilde\wavf\in\widetilde V_{-1}$.
	\item[b)] $\{\wavf_{0,k}\}_{k\in\Z}$ and $\{\widetilde\wavf_{0,k}\}_{k\in\Z}$ are biorthogonal.
	\item[c)] $\{\wavf_{0,k}\}_{k\in\Z}$ is a Riesz basis for $W_0$ and $\{\widetilde \wavf_{0,k}\}_{k\in\Z}$ is a Riesz basis for $\widetilde W_0$.
	\item[d)] For all $k,\ell\in\Z$, $\sprod{\wavf_{0,k}}{\widetilde\scalf_{0,\ell}}=0$ and $\sprod{\widetilde \wavf_{0,k}}{\scalf_{0,\ell}}=0$.
	\item[e)] For all $f\in \CCR$, $\QQ_0f\in W_0$.
\end{enumerate}
Directly from the definition of $\PP_i$ and $\QQ_i$ we obtain that, for all $i,i_0\in\Z$ such that $i\le i_0$, 
the following  holds
\begin{equation}\label{wav:eq:wav_exp}
\PP_if = \sum_{k\in\Z} a_{i_0,k} \scalf_{i_0,k} + \sum_{\ell=i+1}^{i_0}\sum_{k\in\Z} b_{\ell,k}\wavf_{\ell,k}
\end{equation}
with $a_{i_0,k} := \sprod{f}{\widetilde\scalf_{i_0,k}}$ and $b_{\ell,k}:=\sprod{f}{\widetilde \wavf_{\ell,k}}$.

Wavelet theory can be extended to deal with multivariable functions in $L^2(\R^m)$, $m>1$, by considering the tensor product of $m$ GMRAs (see \cite{Daubechies1992}). More precisely, for each $i\in\Z$, we define the subspace  $\bV_i$  as $\bV_i:=V_i\otimes V_i\otimes\cdots\otimes V_i = \cspan\{ \Scalf_{i,\bk}(x),\,\bk\in\Z^m  \}$, in which, with $\bk=(k_1,\dots,k_m)\in\Z^m$,  $\Scalf_{i,\bk}(x):=\scalf_{i,k_1}(x_1)\cdots\scalf_{i,k_m}(x_m)$. Then $(\bV_i)_{i\in\Z}$ forms a GMRA on $L^2(\R^m)$ (i.e. satisfying analogous properties of Definition \ref{def:GMRA}) such that $f\in \bV_i \iff f(2^{i}\cdot)\in \bV_0$, and in which $\Scalf:=\Scalf_{0,0}$ plays the role of a scaling function and the functions $\Scalf_{i,\bk}(x)$, $\bk\in\Z^m$, form a Riesz basis for $\bV_i$. In the same way we construct the dual GMRA $(\widetilde\bV_i)_{i\in\Z}$ and its scaling function $\widetilde\Scalf$. For every $i\in\Z$ and $\bk\in\Z^m$, we construct  $2^m-1$ wavelet functions $\Wavf^h_{i,\bk}$, $h=1,\dots,2^m-1$, by taking all the possible combinations of products of the form
$
g_1(x_1)g_2(x_2)\cdots g_m(x_m)
$, 
with $g_\ell(x_\ell)$ taking the value $\scalf_{i,k_\ell}(x_\ell)$ or $\wavf_{i,k_\ell}(x_\ell)$, except for the case in which $g_\ell(x_\ell)=\scalf_{i,k_\ell}(x_\ell)$ for all $\ell=1,\dots,m$. Namely,
\begin{align*}
\Wavf_{i,\bk}^1(x) &:= \wavf_{i,k_1}(x_1)\scalf_{i,k_2}(x_2)\cdots \scalf_{i,k_m}(x_m)\\
\Wavf_{i,\bk}^2(x)&:=\scalf_{i,k_1}(x_1)\wavf_{i,k_2}(x_2)\scalf_{i,k_3}(x_3)\cdots \scalf_{i,k_m}(x_m)\\
&\cdots\\
\Wavf_{i,\bk}^{2^m-1}(x)&:=\wavf_{i,k_1}(x_1)\wavf_{i,k_2}(x_2)\wavf_{i,k_3}(x_3)\cdots \wavf_{i,k_m}(x_m).
\end{align*}

It can be shown that for every $i,i_0\in\Z$ such that $i\le i_0$, we can express the projection of a function $f\in\cC_c(\R^m)$ into~$\bV_i$ by the following expansion    generalizing \eqref{wav:eq:wav_exp}
\begin{equation}\label{wav:eq:wav_exp_Rd}
\PP_if = \sum_{\bk\in\Z^m} a_{i_0,\bk}\Scalf_{i_0,\bk} + \sum_{h=1}^{2^m-1}\sum_{\ell=i+1}^{i_0}\sum_{\bk\in\Z^m} b_{\ell,\bk}^h\Wavf_{\ell,\bk}^h .
\end{equation}

In the general case, in the wavelet expansion \eqref{wav:eq:wav_exp_Rd} the sum in~$\bk$ ranges over the whole set $\Z^m$. Hence, even for fixed scale~$i$, \eqref{wav:eq:wav_exp_Rd} might consist of infinite terms. Nevertheless, if {\em compactly supported} biorthogonal wavelet and scaling functions are used, the expansion \eqref{wav:eq:wav_exp_Rd} reduces to a {\em finite} sum for each $i$ whenever $f$ has bounded support. 

Some famous families of wavelets with compact support are for instance the {\em Daubechies wavelets} 
\cite{Daubechies1992} or the \emph{B-spline biorthogonal wavelets} of  \cite{Cohen1992}. They mix compact support, smoothness and symmetry.

\bibliographystyle{ieeetran}
\bibliography{biblio}


\end{document}